\pdfoutput=1
\documentclass[a4paper,11pt]{article}
\usepackage[T1]{fontenc}
\usepackage{jcappub}
\usepackage{amsmath,amssymb,graphicx,xcolor,bm}

\newcommand{\oo}{\omega_0}
\newcommand{\wa}{\omega_a}
\newcommand{\wpp}{\omega_p}

\begin{document}

\title{\boldmath Controlled Tension Forecasting: Quantifying Cross-Probe Biases in $\omega_0\omega_a$CDM}

\arxivnumber{2512.04130}

\author[a]{Seokcheon Lee\note{Corresponding author. ORCID: 0000-0003-0861-1300}}
\affiliation[a]{Department of Physics, Institute for Basic Science, Sungkyunkwan University, Suwon 16419, Korea}
\emailAdd{skylee@skku.edu}

\abstract{
Recent analyses combining DESI DR2 BAO, Planck CMB, and Pantheon$+$ SNe have reported mild deviations from the $\Lambda$CDM model. A central challenge is to determine whether these deviations reflect genuine dark-energy evolution or instead arise from cross-probe inconsistencies, prior choices, or mismatches in likelihood construction. Previous work demonstrated that imposing a biased supernova-motivated prior on $\Omega_{m0}$ can artificially displace the BAO-inferred $(\omega_0,\omega_a)$ values from the $\Lambda$CDM expectation. A complementary pedagogic study further showed that differing degeneracy geometries among BAO, CMB, and SNe can generate apparent dark-energy evolution even when the underlying cosmology is exactly $\Lambda$CDM.  Here we present a controlled tension-injection framework designed as a simplified mock-based diagnostic tool for studying how selected probe-level inconsistencies propagate into inferred dark-energy parameters. Self-consistent BAO, CMB, and SNe mock datasets are augmented with parameterized shifts in $(\Omega_{m0}, H_0)$, supernova absolute calibration, and the BAO sound-horizon scale $r_d$. The resulting datasets are analyzed through a unified MCMC pipeline, enabling a direct assessment of how these controlled tensions propagate into biases in $(\omega_0,\omega_a)$ and the pivot equation-of-state parameter $\omega_p$. The results should be interpreted within the restricted setup adopted here: the late-time sector is described in the CPL parametrization and the CMB is represented through compressed distance priors. In this sense, the framework is intended primarily as an illustrative and diagnostic device for identifying probe combinations and degeneracy directions that are more vulnerable to tension-induced dynamical-dark-energy-like shifts, rather than as a parametrization-independent or fully realistic prediction tool.}


\maketitle

\section{Introduction}
\label{sec:intro}

The question of whether dark energy (DE) deviates from a cosmological constant remains central to modern cosmology. Recent analyses of DESI\,DR2 baryon acoustic oscillations (BAO)~\cite{DESI:2024mwx,DESI:2025zgx,DESI:2025fii}, Pantheon$+$ supernovae (SNe)~\cite{Brownsberger:2021uue,Scolnic:2021amr,Brout:2021mpj,Brout:2022vxf,Lane:2023ndt,DES:2025tir}, and Planck cosmic microwave background (CMB) compressed distance priors~\cite{Planck:2013pxb,Planck:2018vyg,Chen:2018dbv,Zhai:2019nad,Lemos:2023xhs} have reported mild indications that $\oo\wa$CDM provides a slightly more flexible phenomenological description than $\Lambda$CDM when heterogeneous probes are combined~\cite{Gialamas:2024lyw,RoyChoudhury:2024wri,Lu:2025gki,Dinda:2025svh,Scherer:2025esj,Sabogal:2025jbo,Gialamas:2025pwv,Dhawan:2025mer,Silva:2025twg,Ishak:2025cay,Wang:2025znm,RoyChoudhury:2025iis}. Some analyses have interpreted these trends as evidence for dynamical dark energy (DDE)~\cite{Giare:2024gpk,Notari:2024zmi,DES:2025bxy}, although such conclusions generally depend sensitively on prior choices, assumptions regarding the sound-horizon scale $r_d$, and the internal consistency of the individual probes~\cite{Ong:2025utx,Giare:2025pzu,Shlivko:2024llw}.

A substantial body of work has emphasized that cross-probe tensions---including the well-known $H_0$ discrepancy between CMB and distance-ladder measurements, as well as inconsistencies between BAO and SNe---or hidden systematics can mimic DDE-like signatures~\cite{Raveri:2018wln,Smith:2022hwi,Colgain:2022nlb,Poulin:2023lkg,Efstathiou:2024xcq,Colgain:2024mtg,CosmoVerseNetwork:2025alb,Lee:2025kbn}. Notably, DES supernova analyses have shown that low-redshift calibration systematics or anchor choices can bias $\Omega_{m0}$ and induce apparent DDE-like behavior~\cite{DES:2024fdw} when analyzed using the Chevallier--Polarski--Linder (CPL) parametrization~\cite{Chevallier:2000qy,Linder:2002et}. This issue has been sharpened by studies demonstrating that the DESI DR1/DR2 claims for evolving DE may be significantly biased by low-redshift SNe systematics~\cite{Huang:2025som}. A recent Bayesian analysis further found that much of the reported preference for $\oo\wa$CDM arises not from true sensitivity to DE evolution, but from the ability of $(\oo,\,\wa)$ to absorb the $\sim3\sigma$ tension between DESI DR2 BAO and DES-Y5 SNe within the $\Lambda$CDM framework~\cite{Notari:2024zmi}.

A complementary pedagogic CPL null test~\cite{Lee:2025grb} supported this point by constructing self-consistent DESI-like BAO, Planck-like CMB, and Pantheon$+$-like SNe mocks from a single fiducial $\Lambda$CDM cosmology. Even in the absence of injected tensions, these probes exhibit distinct degeneracy geometries in the $(\oo,\wa)$ plane~\cite{Clarkson:2010uz,Handley:2019wlz}. When analyzed individually, the probe-dependent ridge structures can induce mild shifts that resemble DDE-like trends while remaining compatible with the underlying $\Lambda$CDM cosmology, similar to patterns seen in real data. Only the full BAO+CMB+SNe combination recovers the fiducial parameters, highlighting the importance of multi-probe coherence for DE inference~\cite{Heavens:2017hkr,DiValentino:2021izs}.

These developments motivate a systematic method to understand how cross-probe inconsistencies affect inferred DE properties. While consistency tests and bias-propagation methods have long been used to assess such effects~\cite{Amara:2007as,MacCrann:2014wfa}, increasingly precise observations now introduce subtler calibration differences, low-redshift anchors, and $r_d$ dependencies, making it necessary to develop a transparent framework that quantifies how such mismatches propagate into shifts of $(\oo,\wa,\wpp)$ within a common inference pipeline~\cite{Tang:2024lmo,Steinhardt:2025znn,Lee:2025kbn,Lee:2025rmg}. In particular, recent assessments of observational constraints have emphasized that interpreting DDE requires care~\cite{Shlivko:2024llw,Giare:2025pzu}, while the broader theoretical context of CPL-based forecasting and the role of pivot formulations has been extensively studied since the DETF report~\cite{Albrecht:2006um}, with further methodological refinements provided in Refs.~\cite{Huterer:2004ch,Martin:2006vv,Scovacricchi:2012fre}.

Beyond these methodological considerations, the parametrization itself can introduce additional structure. Several recent studies have highlighted limitations of the CPL parametrization when applied to heterogeneous or partially inconsistent datasets. These include its sensitivity to basis choices, degeneracy-geometry distortions, and prior-induced artifacts~\cite{Lee:2025ysg,Zhang:2018glx,DES:2024xij,Afroz:2025iwo,Duangchan:2025uzj}. These observations do not by themselves invalidate CPL-based analyses, but they do suggest that apparent departures from $\Lambda$CDM in the $(\oo,\wa)$ plane should be interpreted with caution, especially when the underlying probe combination is not fully coherent.

The purpose of this work is limited and specific. We develop a controlled tension-injection framework, within the CPL parametrization and using a compressed CMB representation, to quantify how selected probe-level inconsistencies propagate into shifts of $(\oo,\wa,\wpp)$ and related cosmological parameters. Our aim is not to reproduce the full complexity of current end-to-end likelihood analyses, but rather to provide a transparent mock-based setting in which the geometric response to controlled inconsistencies can be isolated and studied directly. Therefore, the resulting forecasts should be read as controlled response diagnostics, not as precision surrogates for full real-data analyses.

To this end, we introduce parameterized mismatches in quantities such as $\Omega_{m0}$, $H_0$, the SNe absolute magnitude, redshift-dependent SNe tilts, and the BAO sound horizon~$r_d$, while keeping all other aspects of the mock construction internally consistent. These controlled mismatches are propagated through a unified MCMC engine applied to BAO, CMB, SNe, and their combinations. By comparing the resulting posteriors to a tension-free baseline, we quantify how specific cross-probe inconsistencies appear as shifts in $(\oo,\wa,\wpp)$ and in other derived parameters within this simplified forecasting setup.

In this sense, the present work extends the prior-bias analysis of Ref.~\cite{Lee:2025kbn} and the degeneracy-geometry null tests of Ref.~\cite{Lee:2025grb} into a common controlled framework. The results should be interpreted primarily as diagnostic statements about tension-induced parameter projections in this reduced setting, rather than as parametrization-independent or full-likelihood statements about current data analyses.

To help place the controlled injections in context, we stress that the tension amplitudes explored in this work are not intended as precision reconstructions of any single real-data combination. Rather, they are chosen to span a phenomenologically motivated range broadly comparable to the mild-to-moderate inconsistencies currently discussed in BAO--CMB--SNe analyses, including shifts in $\Omega_{m0}$, $H_0$, SNe calibration, and effective BAO scale assumptions. Therefore, the purpose is not to claim a one-to-one mapping to a specific observational anomaly, but to study, within a transparent mock setting, how inconsistencies of roughly the observed order can be projected into the $(\omega_0,\omega_a,\omega_p)$ sector.

Finally, the remainder of this paper is organized as follows. In Sec.~\ref{sec:method}, we describe the methodological foundations of our framework, including the construction of self-consistent $\Lambda$CDM mocks, the tension-free baseline analysis, and the unified Markov Chain Monte Carlo (MCMC) pipeline used for all probe combinations. Section~\ref{sec:forecast_framework} introduces the parameterized tension models, detailing the individual mismatch channels we consider---such as shifts in $\Omega_{m0}$, $H_0$, the SNe absolute calibration, redshift-dependent SNe tilts, and the BAO sound-horizon scale $r_d$---along with the hybrid and probe-specific tension-injection schemes. We present the main results of the controlled tension-injection runs in Sec.~\ref{sec:run_tension}. We quantify how the injected mismatches propagate into apparent shifts in $(\oo,\wa,\wpp)$ across BAO-only, CMB-only, SNe-only, and all joint-probe configurations, thereby identifying which channels and probe pairings are most susceptible to spurious indications of DDE. Section~\ref{sec:tension_bias_fits} provides a complementary, empirical characterization of these effects by constructing tension--bias transfer functions that map specific injected inconsistencies to the resulting posterior distortions. In Sec.~\ref{sec:discussion}, we discuss the interpretation and limitations of these results, including the dependence on the CPL parametrization, the use of compressed CMB priors, and the extent to which the controlled injections capture realistic sources of cross-probe inconsistency. We also comment on the methodological implications of these issues for future multi-probe analyses involving DESI, Pantheon$+$, Rubin/Legacy Survey of Space and Time (LSST)~\cite{LSSTDarkEnergyScience:2012kar}, and forthcoming CMB experiments such as LiteBIRD~\cite{LiteBIRD:2022cnt} and the Simons Observatory~\cite{SimonsObservatory:2018koc}. Finally, Sec.~\ref{sec:conclusion} summarizes our conclusions and outlines directions for future applications, including survey-optimized tension modeling, pipeline stress testing, and integration with fuller likelihood analyses.

\section{Methodology}
\label{sec:method}

In this section, we summarize the methodological backbone of our controlled tension–injection framework.  We first review the key lessons from previous analysis (hereafter Paper~I) and from the pedagogic CPL null-test project 
(hereafter Paper~II), which jointly motivate the forecasting framework developed in this paper.  In what follows, we focus on the conceptual and mathematical structure of the analysis, with implementation-level details kept implicit to highlight the essential methodology.
\subsection{Connection to Paper~I}
\label{subsec:staf1890_method}

The starting point for the present forecasting framework is Ref.~\cite{Lee:2025kbn} (Paper~I), which investigated how biased external priors on $\Omega_{m0}$ can distort the DE constraints inferred from DESI DR2 BAO data.  In that work, we considered a fiducial flat $\Lambda$CDM cosmology
\begin{equation}
  (\Omega_{m0}^{\rm fid},\,H_0^{\rm fid},\,\oo^{\rm fid},\,\wa^{\rm fid})
  = (0.30,\,70~\mathrm{km s^{-1}Mpc^{-1}},\,-1,\,0), \label{fidCP}
\end{equation}
and analyzed DESI-like BAO distance-ratio observables ($D_M(z)/r_d$ and $D_H(z)/r_d$) using the full tracer-resolved covariance matrix from DESI DR2~\cite{DESI:2025zgx}. The central methodological element of Paper~I was to impose a Gaussian 
prior on $\Omega_{m0}$,
\begin{equation}
  \pi(\Omega_{m0})
  \propto
  \exp\!\left[
    -\frac{(\Omega_{m0}-\Omega_{m0}^{\rm prior})^2}{2\sigma_{\Omega_m}^2}
  \right],
\end{equation}
with $\Omega_{m0}^{\rm prior}$ motivated by a Pantheon$+$-like SNe analysis and deliberately displaced relative to the adopted fiducial value $\Omega_{m0}^{\rm fid}=0.30$.  This setup allowed us to quantify how a mismatched SNe-based prior on 
$\Omega_{m0}$ biases the BAO-inferred CPL parameters, even when the underlying cosmology is exactly $\Lambda$CDM.

Three methodological lessons from Paper~I motivate the present work.  First, the BAO likelihood alone defines a degeneracy ridge in $(\oo,\wa)$ that is sensitive to assumptions about $\Omega_{m0}$.  Second, even a moderate prior offset, $\Delta\Omega_{m0}\equiv\Omega_{m0}^{\rm prior}-\Omega_{m0}^{\rm fid}$, can shift $(\oo,\wa)$ enough to resemble mild DDE-like behaviour.  Third, such shifts can arise from the structure of a partial likelihood---in this case, BAO combined with an external prior---rather than from genuine inconsistencies among independent datasets.  The present framework extends this logic to controlled cross-probe tensions among BAO, CMB, and SNe in a common mock-based setting.

\subsection{Pedagogic CPL Null Test as a Baseline}
\label{subsec:pedagogic_method}

As a complementary foundation, Paper~II~\cite{Lee:2025grb} constructed self-consistent DESI-like BAO distance ratios, Planck-like CMB distance priors, and Pantheon$+$-like SNe mocks from a shared fiducial $\Lambda$CDM cosmology. In contrast to Paper~I, no external priors were imposed and no cross-probe tensions were introduced. The purpose was to examine how much of the apparent DDE-like behaviour seen in real-data analyses could arise from intrinsic likelihood geometry, from the differing degeneracy directions probed by each dataset, and from the complementarity of multi-probe combinations.

Paper~II showed that each probe occupies a distinct degeneracy direction in the $(\Omega_{m0}, H_0,\oo,\wa)$ parameter space, producing mild shifts in $(\oo,\wa)$ when analyzed separately. These shifts arise not from injected tensions or explicit systematics, but from the fact that BAO, CMB, and SNe probe different redshift ranges and different combinations of cosmological parameters. In that sense, they are geometric rather than physical in origin. When the three probes are combined coherently, the fiducial $\Lambda$CDM parameters are recovered, providing a tension-free baseline for the present framework. This baseline allows the shifts discussed in later sections to be interpreted relative to a controlled mock setting without imposed cross-probe inconsistencies.

\subsection{Mock Generation}
\label{subsec:mock_generation}

Starting from the fiducial cosmology in Eq.~\eqref{fidCP}, we construct the baseline mock observables used throughout the analysis.  The goal is to reproduce the statistical properties and covariance structure of the surveys, rather than to emphasize software-specific implementation details. For the BAO sector, we compute the distance ratios $D_M(z)/r_d$ and $D_H(z)/r_d$ at the effective redshifts of the DESI tracers, using covariance matrices matched to DESI DR2.   For the CMB distance prior, we construct the compressed vector $\{R,\ell_A,\omega_b\}$ with means and covariances calibrated to Planck-2018--like constraints, where $R$ is the CMB shift parameter and $\ell_A$ the acoustic angular scale. The SNe sector uses a Pantheon$+$-like Hubble-diagram dataset with full statistical and systematic covariance, with $M$ marginalized as a standard nuisance parameter and optional redshift-dependent tilts $\Delta M(z)$ included in tension-injection tests.

Mock data vectors are drawn from multivariate normal distributions,
\begin{equation}
  \mathbf{d}_{\rm mock}
    = \mathbf{d}_{\rm th}(\theta_{\rm fid})
    + L\,\mathbf{g},
\end{equation}
where $L$ is the Cholesky factor of the survey covariance and $\mathbf{g}$ a vector of standard normal deviates~\cite{Tegmark:1996qs,Morrison:2013tqa,Wadekar:2020hax}.  This construction defines the baseline mock dataset used throughout the analysis. Controlled tensions are introduced later either by modifying selected components of the synthetic data or by altering the parameter assumptions entering the likelihood.

\subsection{BAO, CMB, and SNe Likelihood Structure}
\label{subsec:likelihood_structure}

All probes are analyzed in a unified CPL parameter space,
\begin{equation}
  \theta = \{
  \underbrace{\Omega_{m0}, H_0,\oo,\wa,r_d}_{\text{cosmological}},
  \underbrace{M,\Delta M(z)}_{\text{SNe nuisance}},
  \underbrace{\delta r_d,\delta H_0,\delta\Omega_{m0}}_{\text{tension parameters}}
\},
\end{equation}
where $(\Omega_{m0},H_0,\oo,\wa)$ are the primary cosmological parameters; $r_d$ is the BAO sound horizon; $M$ is the SNe absolute magnitude; $\Delta M(z)$ denotes optional redshift-dependent SNe distortions; and $(\delta r_d,\delta H_0,\delta\Omega_{m0})$ represent the controlled offset parameters that encode the injected cross-probe inconsistencies introduced in Sec.~\ref{sec:forecast_framework}.  Only a subset of these parameters is active in any given 
analysis, with the remainder fixed or set to zero in the tension-free baseline.

In the present work, the late-time response is represented within the CPL basis. Accordingly, the inferred mappings between injected tensions and shifts in $(\oo,\wa)$ should be interpreted as basis-dependent diagnostics within this parametrization, rather than as parametrization-independent response laws.

Each likelihood is assumed Gaussian in the data vector,
\begin{equation}
  -2\ln\mathcal{L}_X(\theta)
  =
  \bigl[
    \mathbf{d}_X^{\rm obs}
    -\mathbf{d}_X^{\rm th}(\theta)
  \bigr]^{\!\mathsf{T}}
  C_X^{-1}
  \bigl[
    \mathbf{d}_X^{\rm obs}
    -\mathbf{d}_X^{\rm th}(\theta)
  \bigr],
\end{equation}
with BAO using $\{D_M/r_d,\,D_H/r_d\}$, CMB using $\{R,\ell_A,\omega_b\}$, and SNe using $\mu(z)$ with optional $\Delta M(z)$.  
Joint analyses use the product likelihood under the assumption that probe-specific covariances adequately capture internal systematics.  In our framework, the tension-injection parameters modify only the theoretical predictions, leaving the baseline mocks unchanged and ensuring that any shift in the recovered posteriors originates solely from the injected inconsistency.

Therefore, the CMB sector is treated in a deliberately reduced form, in which the compressed variables $\{R,\ell_A,\omega_b\}$ are used to isolate the geometric contribution of the CMB to late-time inference. This representation is useful for a controlled forecasting exercise, but it is not equivalent to a full CMB likelihood. In particular, parameters such as $n_s$, $A_s$, and $\tau$, together with their correlations with the geometric sector, are not explicitly restored here, so the resulting degeneracy structure should be regarded as an approximation to the full problem.

Likewise, the Gaussian form assumed for the reduced data vector should be understood as a local approximation in the compressed parameter space. For small to moderate displacements around the fiducial model, this form is adequate for tracking the leading degeneracy directions and the associated bias projections. For larger excursions, however, non-Gaussian features of the reduced CMB likelihood may affect both the detailed posterior shape and the quantitative mapping between injected tensions and inferred shifts in $(\oo,\wa)$.

A related caveat concerns possible non-Gaussianity in the reduced CMB prior itself. Even if the compressed variables $(R,\ell_A,\omega_b)$ provide a useful geometric summary near the fiducial region, the corresponding likelihood need not remain exactly Gaussian once one moves away from that region or projects onto extended late-time parameter directions. In practice, such non-Gaussianity could modify the widths, orientations, and mild asymmetries of the inferred degeneracy structure, and therefore alter the detailed quantitative mapping between injected tensions and posterior shifts. We do not attempt to model that effect here. Accordingly, the CMB sector in the present work should be interpreted as a controlled Gaussianized approximation, suitable for diagnostic forecasting but not as a substitute for a full non-compressed CMB likelihood.

\subsection{Methodological Pipeline and Modular Structure}
\label{subsec:pipeline_scripts}

The full methodology is implemented using a modular analysis pipeline that separates cosmological calculations, mock generation, likelihood evaluation, and sampling.  Cosmological distances, Hubble functions, and BAO ratios are computed within a common CPL module; mock datasets are generated using survey-covariance infrastructure; and BAO, CMB, and SNe likelihoods share a unified parameter interface.  Sampling is performed using the same MCMC framework across all probe combinations.  
This organization ensures that the tension-free baseline and the tension-injected analyses differ only through the controlled offsets applied to the theoretical predictions, which helps isolate the resulting parameter shifts. No changes are made to the mock data vectors or covariances, so that statistical fluctuations and injected systematic mismatches remain distinct within the adopted setup.

\section{Forecasting Framework}
\label{sec:forecast_framework}

This section defines the ingredients of the controlled tension--injection framework used in the present analysis. We first define the tension parameters used to characterize probe-specific inconsistencies in cosmological inference.  We then describe the controlled injection scheme by which these inconsistencies are implemented in otherwise self-consistent BAO, CMB, and SNe mock datasets.  Finally, we present the unified MCMC and Fisher-analysis pipeline through which the injected tensions propagate into posterior distortions of $(\oo,\wa,\wpp)$ and related cosmological parameters.

\subsection{Tension Parameters}
\label{subsec:tension_params}

A central element of this framework is to represent cross-probe inconsistencies through explicit control parameters. To this end, we introduce the tension vector
\begin{equation}
  \Delta\mathbf{t}
  =
  \Bigl(
    \Delta\Omega_{m0},\;
    \Delta H_0,\;
    \Delta M,\;
    \Delta M(z),\;
    \Delta r_d,\;
    \Delta d_{\rm BAO}(z),\;
    \Delta\mu(z)
  \Bigr),
\end{equation}
where each component represents a specified mismatch channel that can be selectively applied to BAO, CMB, or SNe predictions.  
Throughout this paper, the mock catalogs themselves are kept internally consistent; the tension parameters modify only the model predictions $\mathbf{d}_X^{\rm th}(\theta)$.

The full tension vector includes matter-density--like offsets $(\Delta\Omega_{m0},\Delta H_0)$, SNe calibration shifts $(\Delta M,\Delta M(z))$, BAO-scale distortions $(\Delta r_d,\Delta d_{\rm BAO}(z))$, and optional SNe-distance perturbations $\Delta\mu(z)$. 
For clarity, the dominant components used in our controlled experiments are:
\begin{itemize}
\item $\Delta\Omega_{m0}$ (BAO--CMB--SNe matter-density tension),
\item $\Delta H_0$ (CMB--SNe Hubble tension),
\item $\Delta M$ (SNe zero-point calibration offset),
\item $\Delta r_d$ (BAO sound-horizon offset).
\end{itemize}

For the concrete tension--injection experiments discussed in Sec.~\ref{sec:run_tension}, we restrict attention to the reduced tension vector
\begin{equation}
  \Delta\mathbf{t}
  =
  \bigl(
    \Delta\Omega_{m0},\;
    \Delta H_0,\;
    \Delta M,\;
    \Delta r_d
  \bigr),
\end{equation}
which captures the dominant mismatch channels driving cross-probe inconsistencies in DESI--Planck--Pantheon$+$.  Within the present setup, these four components are sufficient to reproduce the leading posterior shifts in $(\oo,\wa,\wpp)$.

For clarity and reproducibility, we summarize the mapping between these tension components and the specific tension-injection runs (runs~2--7).  
These runs constitute the controlled experiments whose results are presented in Sec.~\ref{sec:run_tension}.
\begin{itemize}
  \item \textbf{run2}: SNe-driven $H_0$ tension from a controlled zero-point shift ($\Delta M\neq 0$).
  \item \textbf{run3}: BAO matter-density tension from $\Delta\Omega_{m0}\neq 0$.
  \item \textbf{run4}: CMB sound-horizon tension from $\Delta r_d\neq 0$.
  \item \textbf{run5}: BAO-CMB opposite-sign tension (combined offsets in $\Delta\Omega_{m0}$ and $\Delta r_d$).
  \item \textbf{run6}: SNe anchor--miscalibration scenario with $(\Delta M\neq 0)$ inducing an effective $\Delta H_0$.
  \item \textbf{run7}: Combined multi-probe tension using simultaneous offsets in $(\Delta\Omega_{m0},\Delta H_0,\Delta M,\Delta r_d)$.
\end{itemize}
This mapping summarizes how the formal tension parameters are implemented in the specific forecasting runs considered below.

\subsection{Controlled Tension Injection Scheme}
\label{subsec:tension_injection}

Given a specified tension vector $\Delta\mathbf{t}$, we introduce probe-specific inconsistencies in a controlled manner.  
Three injection modes are considered.

\subsubsection{Data-level injections}

The theoretical predictions entering the mock generation are modified prior to adding survey noise
\begin{equation}
  \mathbf{d}_X^{\rm obs} = \mathbf{d}_X^{\rm th}(\theta_{\rm fid}) + \Delta\mathbf{d}_X(\Delta\mathbf{t}) + L_X\,\mathbf{g}. \label{dXobs}
\end{equation}
Examples include SNe zero-point shifts ($M\to M+\Delta M$), BAO-scale distortions ($r_d\to r_d+\Delta r_d$), or CMB distance-prior shifts modifying $(R,\ell_A)$ to emulate an $H_0$ mismatch.  Within the present framework, these correspond to controlled modifications of the effective probe-level calibration or geometry.

\subsubsection{Prior-level injections}

The mock data remain tension-free, but the likelihood analysis uses a biased external prior, e.g.
\begin{align}
  \pi(\Omega_{m0}) = \mathcal{N}\!\left(\Omega_{m0}^{\rm fid}+\Delta\Omega_{m0},\; \sigma_{\Omega_m}^2 \right). \label{priort}
\end{align}
This corresponds to the prior-bias mechanism identified in Paper~I.

\subsubsection{Hybrid injections}

Some scenarios of interest are represented more naturally by combining data-level and prior-level offsets. Typical examples include a higher SNe-inferred $H_0$ induced by $\Delta M$ together with CMB priors preferring a lower $H_0$, or BAO distance calibrations shifted by $\Delta r_d$ while CMB keeps the fiducial $r_d$.  Within the present setup, such hybrid schemes provide a simple way to represent multi-probe inconsistencies linked to different physical quantities.

\subsection{MCMC Pipeline for Tension Forecasts}
\label{subsec:mcmc_pipeline_forecast}

Once a tension scenario is imposed, the resulting mock datasets are analyzed using the same unified MCMC pipeline as the tension-free baseline.  This ensures that any differences in the recovered posteriors arise solely from the injected inconsistencies.

The purpose of using a common pipeline throughout is diagnostic: it isolates the response of the inferred late-time parameters to controlled inconsistencies without introducing additional changes in the analysis procedure. The resulting forecasts should be interpreted within the same reduced setting adopted throughout this work, namely a CPL-based parameterization with compressed CMB information and simplified tension channels.

\subsubsection{Likelihood construction}

Each probe enters through the Gaussian likelihood
\begin{align}
  -2\ln\mathcal{L}_X(\theta) = \bigl[ \mathbf{d}_X^{\rm obs} -\mathbf{d}_X^{\rm th}(\theta)
  \bigr]^{\!\mathsf{T}}
  C_X^{-1}
  \bigl[ \mathbf{d}_X^{\rm obs} -\mathbf{d}_X^{\rm th}(\theta) \bigr]. \label{LXtheta}
\end{align}

In particular, the Gaussian form is adopted here as a practical approximation for the controlled forecasting problem. This is sufficient for comparing the tension-free baseline with the injected-tension runs on a common footing, but it does not capture all non-Gaussian features that may arise in fuller likelihood treatments, especially in the compressed CMB sector when the displacement from the fiducial region becomes large.

\subsubsection{Parameter vector and priors}

The sampled cosmological parameter vector is $\theta = (\Omega_{m0},H_0,\oo,\wa,r_d,M)$, with flat priors for baseline analyses and appropriately biased priors for prior-level tension injections.

Within this setup, the inferred response in $(\oo,\wa)$ is calibrated specifically in the CPL basis. Alternative late-time parameterizations would in general modify the detailed mapping between the same injected tensions and the resulting posterior shifts.

\subsubsection{Joint sampling}

Probe combinations (BAO+CMB, BAO+SNe, BAO+CMB+SNe) are sampled by multiplying the individual likelihoods.  
We use \texttt{emcee} with $128$ walkers, a $2000$-step burn-in, and a $5000$-step production run~\cite{Foreman-Mackey:2012any}.  The MCMC output includes marginalized posteriors, median constraints, pivot parameters $(a_p,z_p,w_p)$, and derived summaries of posterior geometry.

The goal of this joint sampling is not to reproduce the full statistical complexity of current end-to-end analyses, but to compare, in a controlled and internally consistent way, how different probe combinations respond to the same prescribed inconsistency pattern.

\subsubsection{Fisher-based bias estimates}

To complement the MCMC forecasts, we compute the Fisher matrix
\begin{align}
  F_{\alpha\beta}
  =
  \left\langle
    \frac{\partial^2\chi^2}{\partial\theta_\alpha\partial\theta_\beta}
  \right\rangle_{\rm fid}, \label{Fishersab}
\end{align}
and derive linear-response bias estimates
\begin{align}
  \Delta\theta_\alpha^{\rm (Fisher)}
  =
  \sum_\beta
    (F^{-1})_{\alpha\beta}
    \left(
      \frac{\partial\chi^2}{\partial\theta_\beta}
    \right)_{\Delta\mathbf{t}}. \label{Deltathetaa}
\end{align}
Comparison with the full MCMC results quantifies the accuracy of the linear-response approximation and provides a 
functional tension--bias mapping for each injection scenario.

These Fisher estimates should be regarded as local response diagnostics around the fiducial model. They are useful for tracking the leading direction of the induced bias, but they are not expected to remain quantitatively accurate once the posterior response becomes strongly nonlinear or when several tension channels act simultaneously.

Together, the tension parameters, controlled injection scheme,  and unified MCMC--Fisher pipeline form a coherent architecture for quantifying how cross-probe inconsistencies propagate into apparent shifts in DE parameters.  Within the restricted framework adopted here, this setup provides a transparent and reproducible way to study such projections. Its broader applicability to alternative dark-energy parameterizations or fuller likelihood treatments should be assessed separately.

\section{Results from the Controlled Tension-Injection Runs}
\label{sec:run_tension}

Before presenting the individual tension-injection experiments, we provide a brief overview of the structure of this section.  Runs~2--7 constitute a suite of controlled tests in which specific components of the tension vector $\Delta\mathbf{t}$ are selectively perturbed to emulate well-motivated cross-probe inconsistencies.  Each run isolates a distinct mismatch channel
(\textit{e.g.}, SNe zero-point shifts, BAO matter-density offsets, or CMB sound-horizon perturbations), allowing us to quantify how these controlled deviations propagate into posterior distortions of $(\oo, \wa, \wpp)$ and related parameters.  Run1 serves as the tension-free reference analysis against which all subsequent tension runs are evaluated.  The following subsections present the results of run1--run7 in increasing order of complexity, highlighting the characteristic bias signatures associated with each injected inconsistency.

\subsection{Run1: Tension-Free Baseline Joint Analysis}
\label{subsec:run1}

Run1 establishes the tension-free baseline against which all subsequent tension-injection experiments (run2--7) are evaluated.  In this configuration, no artificial offsets are introduced into any of the probes, and the BAO, CMB, and SNe mock datasets are generated from the common fiducial flat $\Lambda$CDM cosmology. All tension parameters are set to zero,
$\Delta\Omega_{m0}=\Delta H_0=\Delta M=\Delta r_d=0$, so that the three probes form a mutually consistent dataset with no internal inconsistencies.

The joint BAO+CMB+SNe analysis of this configuration provides an essential null test.  The MCMC sampling yields the following median and $1\sigma$ constraints:
\begin{align}
&\Omega_{m0} = 0.2964 \pm 0.0033, \qquad
H_0 = 69.861 \pm 0.229, \nonumber\\[3pt]
&\oo = -1.0138 \pm 0.0416, \qquad
\wa = 0.1440 \pm 0.1370.  \label{run1CP}
\end{align}
These posteriors lie within $1\sigma$ of the fiducial values $(0.30,70,-1,0)$, and the $(\oo, \wa)$ pair aligns with the standard CPL degeneracy band that passes through $(-1,0)$.  No systematic displacement is observed in any direction, indicating that the 
likelihood geometry and numerical integrations behave precisely as expected in the absence of tension.

The corresponding pivot EoS parameters are
\begin{equation}
a_p = 0.7177, \qquad
z_p = 0.3933, \qquad
\wpp = -0.9738 \pm 0.0147,
\end{equation}
again statistically consistent with the fiducial value $\wpp=-1$.  This confirms that the pipeline faithfully reconstructs $\Lambda$CDM when the input mocks contain no probe-level inconsistencies.

Run1 therefore provides the reference point for all subsequent tension runs.  Because the baseline analysis recovers the fiducial cosmology without bias,  any deviations observed in run2--run7 must be attributed to the intentionally injected probe-specific inconsistencies rather than numerical artifacts.  Moreover, Run1 specifies the intrinsic width and orientation of the CPL 
degeneracy directions in the $(\oo, \wa)$ plane, supplying the geometric framework against which tension-induced shifts can be systematically interpreted.

\begin{figure}[t]
    \centering
    \includegraphics[width=0.78\linewidth]{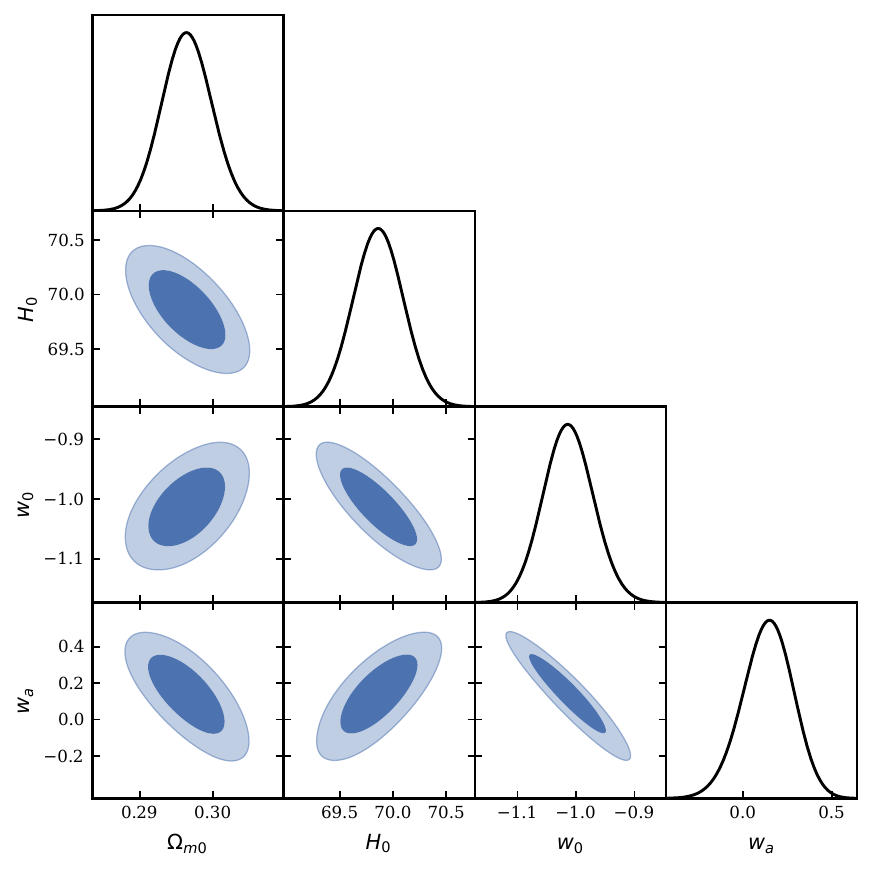}
    \caption{
        \textbf{Run1 (tension-free baseline).}
         Joint posterior for BAO + CMB + SNe without any injected tension.  The recovered parameters are fully consistent with the input $\Lambda$CDM cosmology, demonstrating the internal consistency of the mock datasets and validating the unified likelihood pipeline.}
    \label{fig:run1_triangle}
\end{figure}

Figure~\ref{fig:run1_triangle} displays the joint posterior for $(\Omega_{m0},H_0,\oo, \wa)$.  The contours are centered on the fiducial point and trace the familiar CPL degeneracy direction, with no observable displacement or rotation.  This confirms that Run1 provides a useful baseline against which the subsequent tension-injection experiments can be compared. 

\subsection{Run2: SNe $H_0$ (+3$\sigma$) Tension via Zero-Point Shift}
\label{subsec:run2}

Run2 introduces a controlled tension exclusively in the supernova sector, designed to emulate a SH0ES-like preference for a higher Hubble constant.  This is implemented through a coherent negative shift in the SNe absolute magnitude,$\Delta M = -0.09$, corresponding to a $+3\sigma$ increase in the SNe-inferred Hubble constant ($H_0^{\rm SNe}\!\simeq\!73\,{\rm km\,s^{-1}\,Mpc^{-1}}$)~\cite{Perivolaropoulos:2022khd}.  The BAO and CMB mocks remain fiducial, ensuring that any observed inconsistency in the joint BAO+CMB+SNe posterior originates solely from the SNe calibration bias.

\subsubsection{Posterior Shifts and Degeneracy Geometry}

The joint MCMC constraints obtained in run2 are
\begin{align}
&\Omega_{m0} = 0.3283 \pm 0.0036, \qquad
H_0 = 66.362 \pm 0.210, \nonumber\\[4pt]
&\oo = -0.9096 \pm 0.0392, \qquad
\wa = 0.1507 \pm 0.1150.
\label{run2CP}
\end{align}
Compared with the fiducial cosmology, the joint posterior exhibits a coherent pattern of parameter shifts: the matter density is displaced to higher values, the inferred Hubble constant moves noticeably below the fiducial value, and the CPL parameters migrate toward the region $(\oo>-1,\wa>0)$ typically associated with mild DDE.  

At first sight these trends appear counterintuitive, since the SNe zero-point shift $\Delta M<0$ drives the SNe-only likelihood toward larger $H_0$ and smaller $\Omega_{m0}$.  However, the BAO and CMB likelihoods impose much stronger degeneracy constraints, and these dominate the geometry of the combined posterior.  DESI-like BAO observables favor decreasing $H_0$ in combination with increasing $\Omega_{m0}$ in order to preserve the distance ratios $D_M/r_d$ and $D_H/r_d$, while the CMB acoustic scale $(R,\ell_A)$ remains most stable when a decrease in $H_0$ is accompanied by a shift toward $(\oo>-1,\wa>0)$.  
Because BAO and CMB have significantly greater constraining power than SNe, the joint posterior is pulled along their shared degeneracy direction rather than along the SNe-preferred trend.  This explains why the global fit moves toward $(H_0\downarrow,\Omega_{m0}\uparrow,\oo>-1,\wa>0)$ even though the injected SNe calibration bias alone would have favored the opposite direction.

The qualitative degeneracy directions are summarized in Table~\ref{tab:run2_degeneracy}.

\begin{table}[h!]
\centering
\caption{Qualitative degeneracy directions relevant for run2.  Despite the SNe zero-point tension ($\Delta M<0$) pushing the SNe-only posterior toward higher $H_0$, the much tighter BAO and CMB constraints enforce the opposite trends in the joint analysis.}
\label{tab:run2_degeneracy}
\vspace{0.2cm}
\begin{tabular}{c|c|c}
\hline
\textbf{Probe} & \textbf{Dominant Degeneracy Direction} 
& \textbf{Physical Origin} \\
\hline
SNe 
& $(H_0 \uparrow,\ \Omega_{m0}\downarrow)$ 
& $\Delta M<0$ increases inferred luminosity, raising $H_0$ \\[2pt]
BAO 
& $(H_0 \downarrow,\ \Omega_{m0}\uparrow)$ 
& Preserve $D_M/r_d$ and $D_H/r_d$ ratios \\[2pt]
CMB 
& $(H_0 \downarrow,\ \oo>-1,\ \wa>0)$ 
& Maintain $(R,\ell_A)$ acoustic consistency \\ 
\hline
\end{tabular}
\end{table}

\subsubsection{Tension-Injection Mechanism}

The tension is injected by applying a uniform shift to the SNe distance modulus from $\mu_{\rm th}(z)$ to $\mu_{\rm th}(z) + \Delta M$ with $\Delta M = -0.09$. Since the distance modulus is given by
\begin{align}
\mu = m - M = 5\log_{10}\!\left(\frac{D_L}{\rm Mpc}\right)+25 \label{murun2},
\end{align}
where a negative shift in $M$ corresponds to a brighter absolute magnitude and thus a smaller inferred luminosity distance for a fixed observed magnitude~$m$.  This moves the SNe-only posterior toward a significantly higher Hubble constant.Crucially, the shift is coherent across all SNe, modifying only the global distance scale without affecting the observational covariance matrix.  
BAO and CMB mocks remain unchanged, ensuring that any subsequent inconsistency is a pure SNe–versus–(BAO+CMB) tension.  
The resulting cross-probe conflict forces the joint likelihood to follow the BAO$+$CMB ridge rather than the SNe-preferred direction, thereby producing the net shifts in Eq.~\eqref{run2CP}.

\subsubsection{Pivot Equation-of-State}

The pivot EoS parameters for run2 are
\begin{align}
a_p = 0.6850, \qquad
z_p = 0.4599, \qquad
\wpp = -0.8626 \pm 0.0130.
\label{wprun2}
\end{align}
The deviation of $\omega_p$ from $-1$ indicates that, within this controlled setup, a single-probe calibration bias confined to the SNe distance scale can produce a substantial shift away from the cosmological-constant value. This highlights one of the core lessons of the controlled-tension analysis: posterior deviations from $\Lambda$CDM do not necessarily imply DDE, but may originate from cross-probe inconsistencies.

\subsubsection{Joint Posterior Contours}

\begin{figure}[t]
    \centering
    \includegraphics[width=0.78\linewidth]{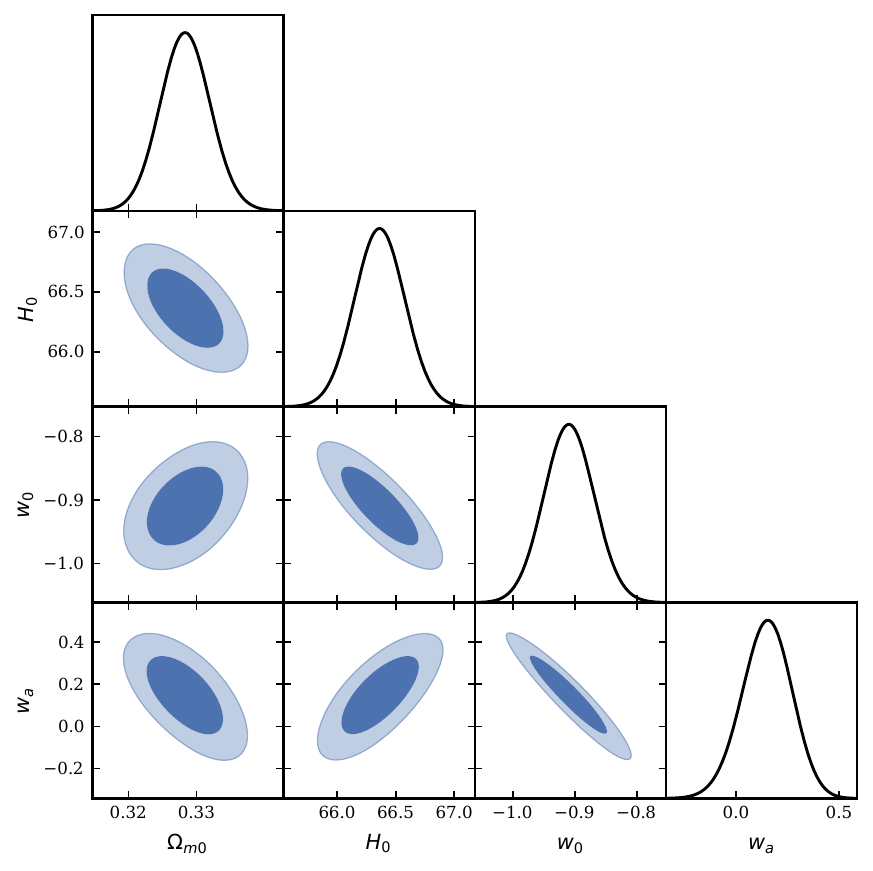}
    \caption{
        \textbf{Run2 (SNe $H_0$ +3$\sigma$ tension).}
        Joint BAO+CMB+SNe posterior when the SNe zero-point is shifted by 
        $\Delta M=-0.09$, mimicking a higher SNe-inferred Hubble constant.  
        BAO and CMB remain fiducial.  
        The contours move along the composite BAO+CMB degeneracy direction toward 
        $(\oo>-1,\ \wa>0)$, showing how, in this controlled setup, a pure SNe calibration bias can project into an apparent dynamical-DE-like shift.}
    \label{fig:run2_triangle}
\end{figure}

Figure~\ref{fig:run2_triangle} shows the full joint posterior for $(\Omega_{m0},H_0,\oo,\wa)$ in run2.  
Relative to the tension-free baseline, the contours shift coherently along the combined BAO+CMB degeneracy direction.  
Although this displacement would normally be interpreted as evidence for DDE, in this case it arises solely from the imposed SNe zero-point miscalibration.

\subsection{Run3: BAO $\Omega_{m0}$ Tension via Distance-Ratio Distortion}
\label{subsec:run3}

In run3 we introduce a controlled tension localized in the BAO sector. This construction is motivated by recent discussions of mild but persistent BAO–CMB inconsistencies in the matter-density sector \cite{Colgain:2024mtg,Chaudhary:2025pcc}, wherein BAO analyses tend to favour slightly lower values of $\Omega_{m0}$ than those inferred from CMB.  This construction mimics a situation in which DESI-like BAO analyses favour a value of $\Omega_{m0}$ that is systematically lower than that preferred by the 
CMB and SNe, while keeping the CMB distance priors and SNe luminosity-distance data fully consistent with the fiducial cosmology.  To achieve this, the BAO observables $D_M/r_d$ and $D_H/r_d$ are distorted along their intrinsic $\Omega_{m0}$-response direction so that the BAO-only likelihood attains its maximum near $\Omega_{m0}^{\rm (BAO)} \simeq 0.28$, two standard deviations below the fiducial value $\Omega_{m0}^{\rm (fid)}=0.30$. No tensions are introduced in the CMB or SNe sectors, ensuring that the only cross-probe inconsistency present in run3 originates from the BAO-preferred matter density.

\subsubsection{Posterior Shifts and Degeneracy Geometry}

The median and $1\sigma$ constraints from the joint BAO+CMB+SNe MCMC analysis are
\begin{align}
&\Omega_{m0} = 0.2874^{+0.0035}_{-0.0034}, \qquad
H_0 = 70.562^{+0.247}_{-0.251}, \nonumber\\[4pt]
&\oo = -1.2012^{+0.0452}_{-0.0442}, \qquad
\wa = 0.7412^{+0.1239}_{-0.1336}. \label{run3CP}
\end{align}
These shifts reflect the characteristic imprint of a BAO-driven low–$\Omega_{m0}$ tension.  The preferred matter density moves slightly below the fiducial value, while $H_0$ shifts upward, lying between the CMB and SNe preferred directions.  Most notably, the CPL parameters migrate into a strongly dynamical region, with $\oo< -1$ and $\wa>0$, resembling a phantom-like form of DDE.  

This behaviour arises from the competition between the BAO-biased ridge direction and the orthogonal degeneracy structure of CMB and SNe.  The BAO distortion pulls the BAO-only likelihood toward $(\Omega_{m0}\!\downarrow, H_0\!\uparrow)$, a direction that also correlates with a tilt in the DE parameters toward $(\oo< -1, \wa>0)$.  CMB distance priors, in contrast, stabilise the acoustic-scale combination $(R,\ell_A)$ most effectively when $H_0$ decreases and $(\oo>-1,\ \wa>0)$,  while SNe prefer a nearly $\Lambda$CDM-like expansion history with $(H_0\uparrow,\Omega_{m0}\downarrow)$ in the presence of a free absolute 
magnitude.  The joint posterior emerges from the compromise between these misaligned degeneracy directions, as summarised in Table~\ref{tab:run3_degeneracy}.

\begin{table}[t]
\centering
\caption{Qualitative degeneracy directions relevant for run3.  The BAO sector is explicitly biased toward a lower matter density, while CMB and SNe remain fiducial.  The joint shifts in Eq.~\eqref{run3CP} arise from the competition among these three misaligned ridge directions.}
\label{tab:run3_degeneracy}
\vspace{0.2cm}
\begin{tabular}{c|c|c}
\hline
\textbf{Probe} & \textbf{Dominant Degeneracy Direction} 
& \textbf{Physical Origin} \\
\hline
BAO (biased) &
$(\Omega_{m0}\downarrow,\ H_0\uparrow,\ \oo< -1,\ \wa>0)$
& Maintain $D_M/r_d$, $D_H/r_d$ under a reduced $\Omega_{m0}$ \\[3pt]
CMB (fiducial) &
$(H_0\downarrow,\ \oo>-1,\ \wa>0)$
& Maintain consistency of $(R,\ell_A)$ \\[3pt]
SNe (fiducial) &
$(H_0\uparrow,\ \Omega_{m0}\downarrow)$
& Shape of $\mu(z)$ with free absolute magnitude \\ 
\hline
\end{tabular}
\end{table}

\subsubsection{Tension-Injection Mechanism}

The BAO tension in run3 is introduced by shifting the BAO-preferred matter density through a coherent distortion of the distance-ratio observables, $d_1(z) \equiv D_M(z)/r_d$ and $d_2(z) \equiv D_H(z)/r_d$, along their response to $\Omega_{m0}$.  
Lowering $\Omega_{m0}$ increases the comoving distance $D_M(z)$ and decreases the expansion rate $H(z)$ at the relevant BAO redshifts, producing correlated shifts in both $D_M/r_d$ and $D_H/r_d$.Within the Fisher approximation, a small perturbation in the BAO-preferred matter density induces
\begin{align}
\Delta d_i(z) \simeq 
\frac{\partial d_i}{\partial \Omega_{m0}}\,
\Delta\Omega_{m0}^{\rm (BAO)}, \label{Deltadi}
\end{align}
and we use this mapping to construct the distorted BAO mock.  
This ensures that the BAO-only posterior shifts along its natural ridge toward lower $\Omega_{m0}$ without altering the internal covariance structure of the survey. The CMB and SNe likelihoods remain unmodified, so the only injected inconsistency is the BAO-preferred value $\Omega_{m0}^{\rm (BAO)}\simeq 0.28$.
When the three probes are combined, the full likelihood
\begin{align}
\mathcal{L}_{\rm tot}
= \mathcal{L}_{\rm BAO}^{(\Omega_{m0}\text{-biased})}\,
  \mathcal{L}_{\rm CMB}^{\rm (fid)}\,
  \mathcal{L}_{\rm SNe}^{\rm (fid)} \label{mathcalLrun3}
\end{align}
must reconcile three misaligned degeneracy directions.  

\subsubsection{Pivot Equation-of-State}

Recasting the CPL parameters in terms of the pivot EoS minimises the covariance between the DE amplitude and its slope.  For run3, we obtain
\begin{align}
a_p = 0.71, \qquad
z_p \simeq 0.41, \qquad
\wpp = -1.045 \pm 0.018. \label{wprun3}
\end{align}
The pivot value lies moderately below $-1$, consistent with the phantom-like shift seen in $(\oo,\wa)$.  For comparison, run2—where the injected tension resided entirely in the SNe sector—produced a pivot value above $-1$, corresponding to a 
quintessence-like trend.  In contrast, run3 shows that shifting the BAO sector drives the pivot EoS below $-1$, illustrating that the direction of the apparent DE deviation is primarily determined by the probe in which the tension resides.

\subsubsection{Joint Posterior Contours}

\begin{figure}[t]
    \centering
    \includegraphics[width=0.78\linewidth]{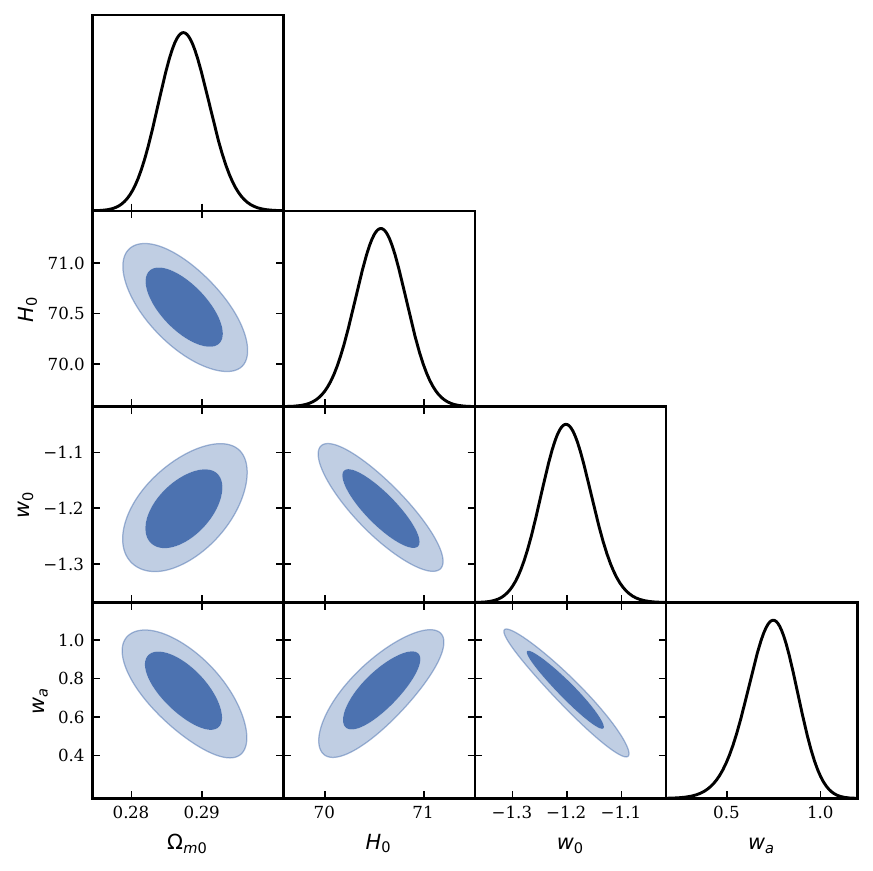}
    \caption{ \textbf{Run3 (BAO $\Omega_{m0}$ tension).} Triangle plot from the joint BAO+CMB+SNe analysis.  The BAO sector is biased toward a lower matter density, while the CMB and SNe mocks remain fiducial.  The resulting posterior is displaced toward 
 $\Omega_{m0}\simeq 0.287$, $H_0\simeq 70.6$, and $(\oo,\wa)\simeq(-1.20,\,0.74)$, producing a pronounced phantom-like shift in the effective CPL parameters, even though the underlying cosmology is exactly $\Lambda$CDM.}
    \label{fig:run3_triangle}
\end{figure}

Figure~\ref{fig:run3_triangle} presents the joint posterior for $(\Omega_{m0},H_0,\oo,\wa)$.  Relative to the tension-free baseline and the SNe-driven tension in run2, the $(\oo,\wa)$ contours show a pronounced shift toward $(\oo < -1,\ \wa > 0)$, consistent with the pivot trend in Eq.~\eqref{wprun3}.  In the $(\Omega_{m0},H_0)$ plane, the BAO-induced preference for lower 
$\Omega_{m0}$ is only partially realised: CMB and SNe pull the matter density back upward toward $0.30$. The compensation occurs primarily in the DDE direction rather than through shifts in $H_0$, since the BAO and CMB degeneracy directions intersect at shallow angles in the $(\oo,\wa)$ plane. This makes the DDE parameters more efficient than $H_0$ in reconciling the cross-probe inconsistency.

\subsection{Run4: CMB Acoustic-Scale Tension via Low–$H_0$ Distance Prior}
\label{subsec:run4}
In run4 we introduce a controlled tension localized in the CMB sector,  implemented as an effective shift in the CMB-preferred Hubble constant. The CMB mock is modified such that the distance-prior combination $(R,\ell_A)$ favours a value of
$H_0^{\rm (CMB)} \simeq 65~{\rm km\,s^{-1}Mpc^{-1}}$ which is approximately $5~{\rm km\,s^{-1}Mpc^{-1}}$ lower than the fiducial $H_0^{\rm (fid)}=70~{\rm km\,s^{-1}Mpc^{-1}}$.  The BAO and SNe mocks remain fully fiducial ($\Delta\Omega_{m0}=\Delta M=\Delta r_d=0$), ensuring that the only cross-probe inconsistency in run4 originates from the CMB side through its low–$H_0$ preference~\cite{Chen:2018dbv,DiValentino:2020zio,Efstathiou:2021ocp}.  This setup mimics a Planck-like acoustic-scale tension: lowering $H_0$ while keeping $r_d$ fixed increases the angular size of the sound horizon ($\theta_s = r_d / D_A$), reproducing the characteristic direction of the CMB–local $H_0$ discrepancy.

\subsubsection{Posterior Shifts and Degeneracy Geometry}

The median and $1\sigma$ constraints obtained from the joint MCMC analysis are
\begin{align}
&\Omega_{m0} = 0.3384 \pm 0.0031, \qquad
H_0 = 69.875 \pm 0.166, \nonumber\\[4pt]
&\oo = -1.1781 \pm 0.0166, \qquad
\wa = 1.0822 \pm 0.0258.
\label{run4CP}
\end{align}
These shifts reflect the characteristic imprint of a CMB-driven low–$H_0$ tension.  The inferred matter density increases well above the fiducial value, the resulting $H_0$ lies slightly below $70$, and the CPL parameters move deeply into the phantom-like region with $\oo< -1$ and $\wa>0$.  The strong excursion in $(\oo,\wa)$ arises from the well-known sensitivity of the acoustic-scale combination $(R,\ell_A)$ to changes in $H_0$:  a lower $H_0$ increases the sound-horizon angle $\theta_s$, which the MCMC partially compensates by shifting toward a late-time expansion history consistent with $\oo< -1$ and $\wa>0$. The multi-probe geometry underlying these shifts is summarised in Table~\ref{tab:run4_degeneracy}.  The biased CMB mock strongly prefers $(H_0\downarrow,\ \oo< -1,\ \wa>0)$,  while the fiducial BAO ridges favour $(H_0\downarrow,\ \Omega_{m0}\uparrow)$, and the SNe degeneracy tends toward $(H_0\uparrow,\ \Omega_{m0}\downarrow)$.  The joint posterior aligns primarily with the BAO+CMB composite ridge, which links a decrease in $H_0$ to simultaneous increases in $\Omega_{m0}$ and a strong tilt toward $(\oo < -1, \wa > 0)$.  As in run3, the inconsistency is accommodated primarily through shifts in the DE parameters, but here the effect is substantially amplified due to the steep sensitivity of $(R,\ell_A)$ to $H_0$. 

\begin{table}[t]
\centering
\caption{Qualitative degeneracy directions relevant for run4.  
Only the CMB mock is biased, favouring a lower $H_0$, while the BAO and SNe 
data remain fiducial.  
The interaction of these ridge directions determines the joint posterior 
structure in Eq.~\eqref{run4CP}.}
\label{tab:run4_degeneracy}
\vspace{0.2cm}
\begin{tabular}{c|c|c}
\hline
\textbf{Probe} & \textbf{Dominant Degeneracy Direction} 
& \textbf{Physical Origin} \\
\hline
CMB (biased) &
$(H_0\downarrow,\ \oo< -1,\ \wa>0)$
& Maintain $(R,\ell_A)$ under reduced $H_0$ \\[3pt]
BAO (fiducial) &
$(H_0\downarrow,\ \Omega_{m0}\uparrow)$
& Stabilise $D_M/r_d$ and $D_H/r_d$ \\[3pt]
SNe (fiducial) &
$(H_0\uparrow,\ \Omega_{m0}\downarrow)$
& Geometry of $\mu(z)$ with free $M$ \\
\hline
\end{tabular}
\end{table}

\subsubsection{Tension-Injection Mechanism}

The CMB-side tension in run4 is introduced by modifying the distance-prior vector $(R,\ell_A)$ so that it is internally consistent with a lower CMB-preferred Hubble parameter,
\begin{align}
H_0^{\rm (CMB)} = H_0^{\rm (fid)} + \Delta H_0^{\rm (CMB)}
                 = 70 - 5
                 \simeq 65~{\rm km\,s^{-1}Mpc^{-1}} . \label{H0run4}
\end{align}
For fixed $(\Omega_{m0},\oo,\wa,r_d)$, a decrease in $H_0$ increases the sound-horizon angular scale $\theta_s \equiv r_s(z_\ast)/D_A(z_\ast)$, and hence shifts both the acoustic peak position $\ell_A \propto \pi/\theta_s$ and the CMB shift parameter $R = \sqrt{\Omega_{m0} H_0^2}\, D_M(z_\ast)$. We implement the bias by mapping the fiducial distance-prior vector
$\mathbf{d}_{\rm CMB}^{\rm (fid)} = (R_{\rm fid}, \ell_{A,{\rm fid}})$ to a shifted vector
\begin{align}
\mathbf{d}_{\rm CMB}^{\rm (biased)}
   = \mathbf{d}_{\rm CMB}^{\rm (fid)}
     + \Delta\mathbf{d}_{\rm CMB}, \label{drun5}
\end{align}
where $\Delta\mathbf{d}_{\rm CMB}$ corresponds to the change induced when
$H_0$ is lowered by $5\,{\rm km\,s^{-1}Mpc^{-1}}$ while keeping all other
fiducial parameters fixed.  
In the Fisher approximation, this perturbation can be written as
\begin{align}
\Delta d_i
   \simeq 
   \frac{\partial d_i}{\partial H_0}\,
   \Delta H_0^{\rm (CMB)},
   \qquad d_i \in \{R,\ell_A\}, \label{drun52}
\end{align}
ensuring that the CMB mock remains internally self-consistent but corresponds to a cosmology with a lower preferred Hubble constant.

No modifications are applied to the BAO or SNe sectors, so the full likelihood
\begin{align}
\mathcal{L}_{\rm tot}
  = \mathcal{L}_{\rm BAO}^{\rm (fid)}
    \mathcal{L}_{\rm CMB}^{(H_0\text{-biased})}
    \mathcal{L}_{\rm SNe}^{\rm (fid)} \label{mathcalLrun4}
\end{align}
contains a single, well-defined inconsistency arising from the CMB acoustic scale.  When combined with the BAO ridge (linking $H_0\downarrow$ to $\Omega_{m0}\uparrow$) and the SNe ridge (linking $H_0\uparrow$ to $\Omega_{m0}\downarrow$),
the biased CMB priors force the joint solution toward the direction $(H_0\downarrow,\ \Omega_{m0}\uparrow,\ \oo< -1,\ \wa>0)$, matching the behaviour observed in Eq.~\eqref{run4CP}.

\subsubsection{Pivot Equation-of-State}

The pivot parameters extracted from the run4 posterior are
\begin{align}
a_p = 1.0806, \qquad
z_p = -0.0746, \qquad
\wpp = -1.2654 \pm 0.0186.
\label{wprun4}
\end{align}
The negative value of \(z_p\) indicates that the pivot direction has rotated past \(z=0\), a clear sign that the CMB-induced tension forces the fit to compensate through rapid late-time evolution of the DE EoS.  The pivot value lies far below $-1$, indicating a pronounced phantom-like behaviour driven almost entirely by the acoustic-scale inconsistency in the CMB sector.  Compared with run3—which exhibited only a mild phantom-like shift—run4 demonstrates the substantially stronger tendency of the CMB acoustic scale to push the posterior toward $\oo < -1$ when $H_0$ is artificially decreased.

\subsubsection{Joint Posterior Contours}

\begin{figure}[t]
    \centering
    \includegraphics[width=0.78\linewidth]{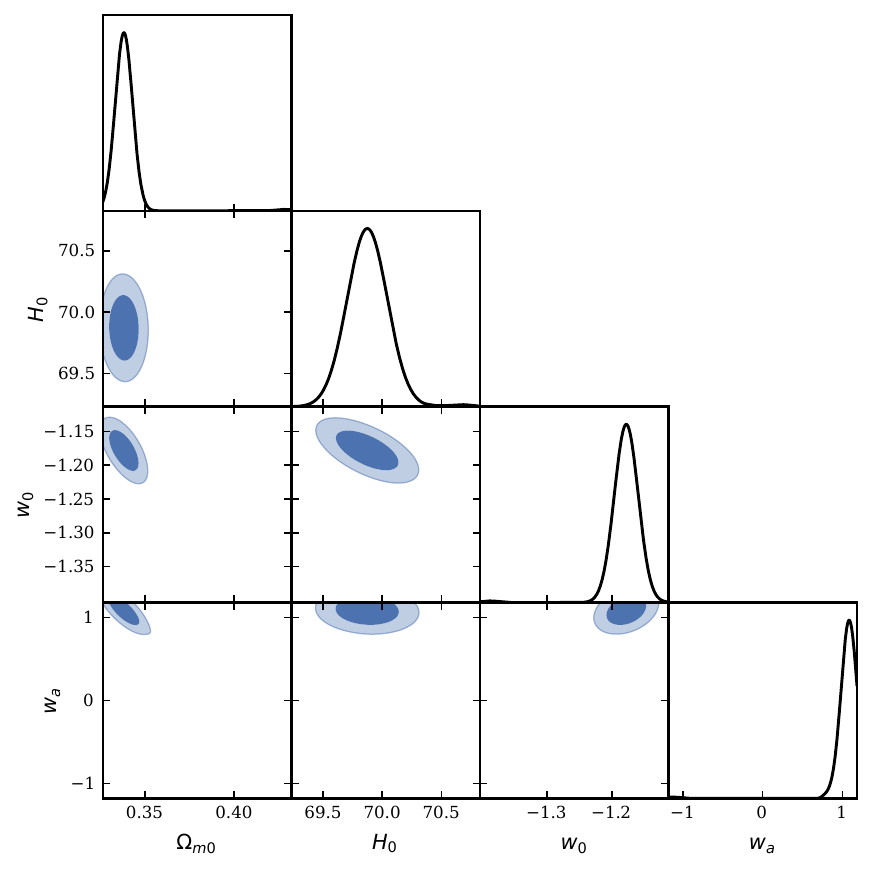}
    \caption{
        \textbf{Run4 (CMB-side $H_0$ / acoustic-scale tension).}
        Joint posterior for BAO+CMB+SNe when only the CMB mock is biased 
        toward lower $H_0$.  
        The resulting displacement toward 
        $\Omega_{m0}\simeq 0.338$, $H_0\simeq 69.9$, and 
        $(\oo,\wa)\simeq(-1.18,\,1.08)$ traces a steep, phantom-like degeneracy direction.  In this controlled setup, the apparent DDE-like displacement arises entirely from the imposed CMB-side tension.}
    \label{fig:run4_triangle}
\end{figure}

Figure~\ref{fig:run4_triangle} presents the joint posterior contours for $(\Omega_{m0},H_0,\oo,\wa)$ obtained in run4.  
The $(\oo,\wa)$ ellipses are displaced far into the phantom-like quadrant,  consistent with the pivot trend in Eq.~\eqref{wprun4}, while the $(\Omega_{m0},H_0)$ contours reflect the combined BAO+CMB ridge that links lower $H_0$ with higher $\Omega_{m0}$.  Although the underlying cosmology is exactly $\Lambda$CDM, the imposed CMB-side tension drives the combined posterior well away from the fiducial region in the $(\oo,\wa)$ plane. 

\subsection{Run5: Opposite-Directed BAO--CMB Tensions}
\label{subsec:run5}
Run5 introduces two deliberately opposite-directed tensions in the BAO and CMB sectors~\cite{Bernal:2016gxb,Lopez-Hernandez:2025lbj}.  The BAO mock is distorted so that the BAO-only likelihood prefers a higher matter density,  $\Omega_{m0}^{\rm (BAO)} \simeq 0.33$,
while the CMB distance prior is modified to mimic a lower Hubble constant, $H_0^{\rm (CMB)} \simeq 65$. The SNe dataset remains strictly fiducial.  This configuration generates the strongest mutual inconsistency among the tension scenarios studied here, with BAO and CMB pulling the parameters in nearly orthogonal directions within the $(\Omega_{m0},H_0,\oo,\wa)$ space.  Run5 therefore constitutes a clean test of how conflicting degeneracy directions interfere when SNe are added as a third geometric anchor.

\subsubsection{Posterior Shifts and Degeneracy Geometry}

The median and $1\sigma$ constraints obtained from the BAO+CMB+SNe MCMC analysis are
\begin{align}
&\Omega_{m0} = 0.3371 \pm 0.0031, \qquad
H_0 = 69.611 \pm 0.165, \nonumber\\[4pt]
&\oo = -1.1533 \pm 0.0157, \qquad
\wa = 1.0563 \pm 0.0228.
\label{run5CP}
\end{align}
The combined posterior reflects a coherent compromise between the competing BAO and CMB tension directions.  The matter density shifts upward by roughly $12\%$ relative to the fiducial value, dominated by the BAO-induced pull toward higher $\Omega_{m0}$, while the inferred Hubble constant decreases slightly below $70$, moving toward the CMB-preferred low–$H_0$ direction.  The CPL parameters respond to the combined BAO+CMB geometry by migrating into the $(\oo< -1,\ \wa>0)$ quadrant associated with phantom-like evolution, but the shift is noticeably milder than in run4 because the BAO-side distortion partially counteracts the CMB-induced tilt.

Taken together, the three probes generate a highly non-parallel set of degeneracy directions as summarised in Table~\ref{tab:run5_degeneracy}.  The BAO bias drives the parameters toward $(\Omega_{m0}\uparrow, H_0\downarrow)$, while the CMB bias enforces $(H_0\downarrow,\, \oo< -1,\, \wa>0)$, and SNe remain centred on the fiducial $(\Omega_{m0},H_0)$ ridge.  Because the BAO and CMB tensions pull the solution in nearly orthogonal directions in the $(\Omega_{m0}, H_0, \oo,  \wa)$ space, the resulting posterior corresponds to a vector-like compromise between these incompatible ridges.  This yields the balanced displacement seen in Eq.~\eqref{run5CP}, with a phantom-like shift that is substantially weaker than in run4 due to the partial cancellation of the BAO and CMB tension vectors.

\begin{table}[t]
\centering
\caption{Qualitative degeneracy directions for run5.  
The BAO mock is biased toward high $\Omega_{m0}$, while the CMB mock is biased 
toward low $H_0$.  
The SNe sector remains fiducial.  
The joint posterior reflects a compromise between the nearly orthogonal 
tension directions.}
\label{tab:run5_degeneracy}
\vspace{0.2cm}
\begin{tabular}{c|c|c}
\hline
\textbf{Probe} & \textbf{Dominant Degeneracy Direction} 
& \textbf{Physical Origin} \\
\hline
BAO (biased high $\Omega_{m0}$) &
$(\Omega_{m0}\uparrow,\ H_0\downarrow)$
& Maintain $D_M/r_d$ and $D_H/r_d$ at high $\Omega_{m0}$ \\[3pt]
CMB (biased low $H_0$) &
$(H_0\downarrow,\ \oo< -1,\ \wa>0)$
& Keep $(R,\ell_A)$ fixed under reduced $H_0$ \\[3pt]
SNe (fiducial) &
$(H_0\uparrow,\ \Omega_{m0}\downarrow)$
& Luminosity-distance geometry with free $M$ \\
\hline
\end{tabular}
\end{table}

\subsubsection{Tension-Injection Mechanism}

The BAO tension is implemented through a positive shift $\Delta\Omega_{m0}^{\rm (BAO)}=+0.03$, which modifies both 
$D_M(z)/r_d$ and $D_H(z)/r_d$ along their intrinsic sensitivity to the matter density.  At fixed $(H_0,\oo,\wa,r_d)$, increasing $\Omega_{m0}$ raises $H(z)$ and reduces $D_M(z)$, pulling the BAO-only posterior toward the high–$\Omega_{m0}$ region. Simultaneously, the CMB mock is adjusted via a negative shift in the preferred Hubble parameter,
$\Delta H_0^{\rm (CMB)} = -5~{\rm km\,s^{-1}Mpc^{-1}}$, which distorts the acoustic-scale variables $(R,\ell_A)$ and shifts the CMB posterior toward $(H_0\downarrow,\ \oo< -1,\ \wa>0)$. With SNe left fiducial, the full likelihood
\begin{align}
\mathcal{L}_{\rm tot}
 = \mathcal{L}_{\rm BAO}^{(\Omega_{m0}\textrm{-biased})}
   \mathcal{L}_{\rm CMB}^{(H_0\textrm{-biased})}
   \mathcal{L}_{\rm SNe}^{\rm(fid)} \label{matchcalLrun5}
\end{align}
must reconcile three incompatible geometric ridges.  The resulting solution is the balanced displacement visible in 
Eq.~\eqref{run5CP}, lying between the BAO and CMB preferred directions. Importantly, the BAO-induced shift in $(\Omega_{m0},H_0)$ and the CMB-induced shift in $(H_0,\oo, \wa)$ span nearly orthogonal directions in parameter space.  
This guarantees that the joint-posterior displacement is not dominated by either probe alone, but instead reflects the geometric interference between the two distinct tension vectors.

\subsubsection{Pivot Equation-of-State}

The pivot parameters for run5 are
\begin{align}
a_p = 0.9628, \qquad
z_p = 0.0387, \qquad
\wpp = -1.1143 \pm 0.0151.
\label{wprun5}
\end{align}
The positive value of $z_p$ indicates a pivot scale located slightly above the present epoch.  The pivot EoS remains below $-1$, consistent with the phantom-like shift induced jointly by the BAO and CMB tensions, but the deviation from the fiducial value is significantly less extreme than in run4.  This illustrates how oppositely directed tension vectors can partially cancel, thereby reducing the apparent severity of the dynamical-DE-like shift in the pivot plane, despite the underlying cosmology being exactly $\Lambda$CDM. In this sense, run5 shows that dynamical-DE-like posteriors may be either amplified or suppressed depending on the angular structure of the tension vectors in parameter space. 

\subsubsection{Joint Posterior Contours}

\begin{figure}[t]
    \centering
    \includegraphics[width=0.78\linewidth]{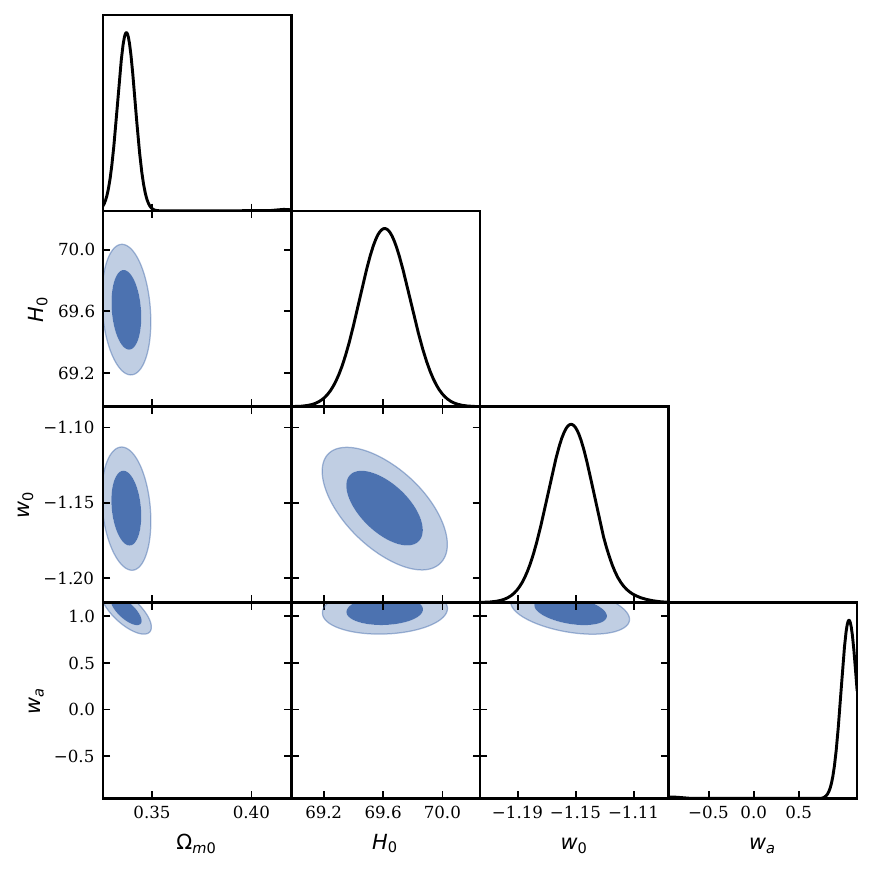}
    \caption{
        \textbf{Run5 (opposite-directed BAO--CMB tensions).}  
        Joint posterior for BAO+CMB+SNe when the BAO mock is biased toward 
        high $\Omega_{m0}$ and the CMB mock toward low $H_0$.  
        The recovered parameters,
        $\Omega_{m0}\simeq 0.337$, $H_0\simeq 69.6$, and 
        $(\oo,\wa)\simeq(-1.15,\,1.06)$,
        reflect a compromise between mutually inconsistent ridge directions.  
        The apparent DDE signature is weaker than in run4 due to the 
        partial cancellation of BAO and CMB shifts.}
    \label{fig:run5_triangle}
\end{figure}

Figure~\ref{fig:run5_triangle} displays the joint posterior contours for $(\Omega_{m0},H_0,\oo,\wa)$ in run5.  The enhancement of $\Omega_{m0}$ and the suppression of $H_0$ produce a characteristic diagonal orientation in the $(\oo,\wa)$ plane, reflecting the interference between BAO and CMB degeneracies.  The elongation in $\wa$ corresponds to the partial cancellation of the tension directions, which broadens the allowed region relative to run4.  

\subsection{Run6: SNe Anchor Miscalibration (Zero-Point Shift)}
\label{subsec:run6}

A coherent shift in the SNe absolute magnitude is strictly degenerate with the Hubble constant~\cite{Efstathiou:2021ocp,Camarena:2021jlr}, yet its ability to bias cosmological parameters is tightly constrained when combined with high-redshift geometric probes via the inverse distance ladder~\cite{BOSS:2014hhw, DES:2018rjw}. Run6 explicitly demonstrates this behaviour within our controlled tension-injection framework. In contrast to run2—which imposed a large, SH0ES-like negative $\Delta M_B$ to enforce a strong high-$H_0$ preference—the perturbation applied here is deliberately subtle. This design isolates the intrinsic geometric sensitivity of the global BAO+CMB+SNe fit to SNe calibration errors, avoiding the strongly non-linear parameter distortions observed in runs~3--5.

\subsubsection{Posterior Shifts and Degeneracy Geometry}

The median and $1\sigma$ constraints from the joint MCMC analysis are
\begin{align}
&\Omega_{m0} = 0.2868 \pm 0.0033, \qquad
H_0 = 71.048 \pm 0.237, \nonumber\\[4pt]
&\oo = -1.0405 \pm 0.0427, \qquad
\wa = 0.1097 \pm 0.1460. \label{run6CP}
\end{align}
These shifts display the characteristic fingerprint of a SNe-only zero-point perturbation.  The inferred Hubble constant moves slightly above the fiducial value,  consistent with the SNe sector favoring a brighter absolute magnitude.  Conversely, the matter density shifts mildly downward, following the correlation direction of the SNe ridge.  The CPL parameters exhibit only a mild tilt toward $(\oo< -1,\ \wa>0)$, and the displacement is substantially smaller than in runs~3--5,  reflecting the much weaker geometrical imprint of a small SNe-only calibration shift. This behaviour reflects the limited geometric leverage of SNe: a perturbation in $\Delta M$ affects the global fit primarily through its degeneracy with $H_0$, and only secondarily through correlations with $(\oo,\wa)$.

The relevant degeneracy directions for each probe are summarised in Table~\ref{tab:run6_degeneracy}.  The SNe distortion pushes the posterior toward $(H_0\uparrow,\ \Omega_{m0}\downarrow)$, while the fiducial BAO and CMB likelihoods favour the opposing trends.  The joint posterior therefore adopts an intermediate solution that remains very close to the fiducial expansion history.

\begin{table}[t]
\centering
\caption{Qualitative degeneracy directions for run6.  
Only the SNe distance scale is perturbed; BAO and CMB remain fiducial.  
The moderate shifts in Eq.~\eqref{run6CP} arise from the competition between 
the SNe ridge and the nearly orthogonal BAO and CMB degeneracies. The SNe ridge corresponds to the near-exact degeneracy between $H_0$ and the 
SNe absolute magnitude, so its shifts primarily manifest as motions along the 
$(H_0,\Omega_{m0})$ plane rather than in $(\oo,\wa)$.
}
\label{tab:run6_degeneracy}
\vspace{0.2cm}
\begin{tabular}{c|c|c}
\hline
\textbf{Probe} & \textbf{Dominant Degeneracy Direction}
& \textbf{Physical Origin} \\
\hline
SNe (miscalibrated) &
$(H_0\uparrow,\ \Omega_{m0}\downarrow)$
& Zero-point shift in $\mu(z)$ \\[3pt]
BAO (fiducial) &
$(H_0\downarrow,\ \Omega_{m0}\uparrow)$
& Preserve $D_M/r_d$ and $D_H/r_d$ \\[3pt]
CMB (fiducial) &
$(H_0\downarrow,\ \oo>-1,\ \wa>0)$
& Acoustic-scale consistency \\ 
\hline
\end{tabular}
\end{table}

\subsubsection{Tension-Injection Mechanism}

The SNe miscalibration is implemented by shifting the theoretical distance modulus from $\mu_{\rm th}(z)$ to $\mu_{\rm th}(z) + \Delta M$ with no changes applied to BAO or CMB.  Since the distance modulus is given by
\begin{align}
\mu = m - M = 5\log_{10}(D_L/{\rm Mpc}) + 25 \label{mumodule} \,,
\end{align}
a brighter absolute magnitude ($\Delta M<0$) corresponds to a smaller inferred distance and hence a larger inferred value of $H_0$ at fixed apparent magnitude.  When the joint likelihood is evaluated, BAO and CMB counteract this trend by pulling toward lower $H_0$, resulting in the balanced shifts observed in Eq.~\eqref{run6CP}.  Because the imposed $\Delta M$ is coherent across all redshifts, the SNe 
likelihood merely slides along its intrinsic $(H_0,\Omega_{m0})$ degeneracy 
without altering its shape.  The substantially higher constraining power of 
BAO and CMB therefore stabilizes the joint posterior, suppressing the impact 
of the SNe-only perturbation.

\subsubsection{Pivot Equation-of-State}

The pivot EoS parameters for run6 are
\begin{align}
a_p = 0.7300, \qquad
z_p = 0.3699, \qquad
\wpp = -1.0115 \pm 0.0153.
\label{wprun6}
\end{align}
This remains essentially indistinguishable from $\wpp = -1$ within the statistical uncertainty, underscoring that a mild SNe zero-point offset has negligible impact on global DE constraints when BAO and CMB data are included.  This sharply contrasts with the phantom-like pivots seen in runs~3--5,  highlighting the significantly lower sensitivity of SNe calibration to the combined BAO+CMB geometry.

\subsubsection{Joint Posterior Contours}

\begin{figure}[t]
    \centering
    \includegraphics[width=0.78\linewidth]{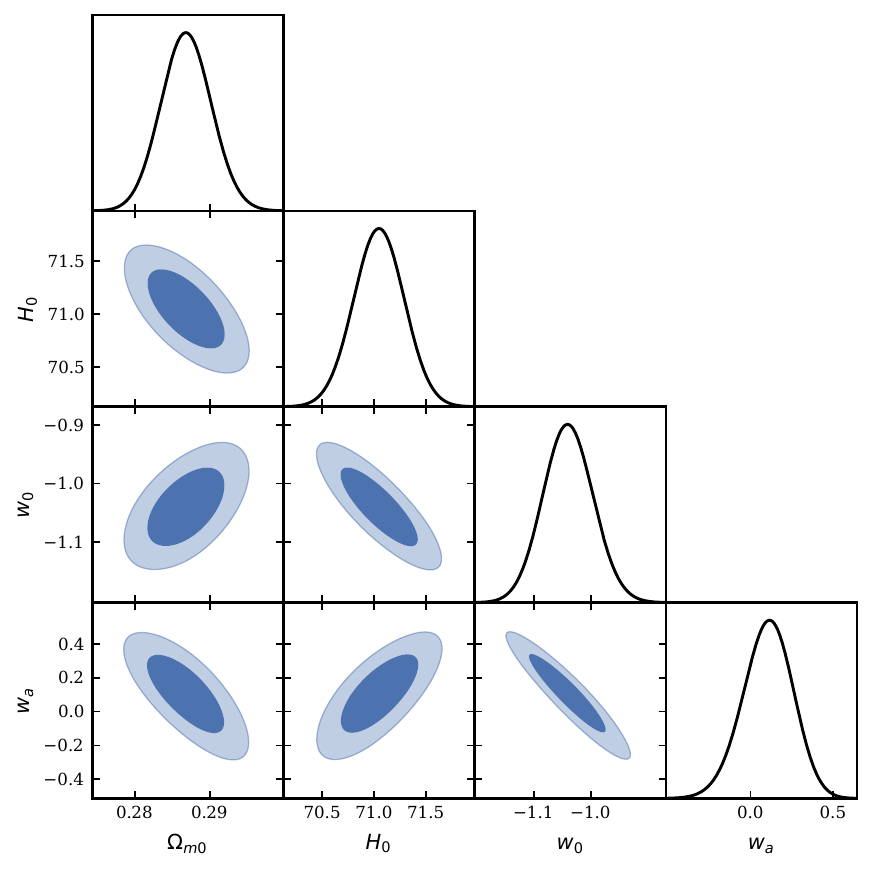}
    \caption{
        \textbf{Run6 (SNe anchor miscalibration).}
        Joint BAO+CMB+SNe posterior when only the SNe zero point is shifted.  
        The recovered parameters,
        $\Omega_{m0}\simeq 0.287$, $H_0\simeq 71.0$, 
        $(\oo,\wa)\simeq(-1.04,\,0.11)$,
        and $\wpp\simeq -1.01$,
        remain close to the fiducial $\Lambda$CDM cosmology.
        This illustrates the comparatively weak leverage of SNe-only 
        miscalibration on the global DE constraints.}
    \label{fig:run6_triangle}
\end{figure}

Figure~\ref{fig:run6_triangle} displays the joint posterior for $(\Omega_{m0},H_0,\oo,\wa)$.  The contours show only modest displacements relative to the tension-free baseline.  Most of the shift is confined to the $(\Omega_{m0},H_0)$ panel, tracing the 
shallow SNe ridge.  BAO and CMB effectively stabilize the global fit, preventing the SNe miscalibration from producing any substantial deviation from $\Lambda$CDM.

\subsection{Run7: Combined Multi-Probe Systematic Tension}
\label{subsec:run7}

Run7 represents the most complex configuration in our controlled-tension suite.  Unlike runs~2--6, which each introduce a single-probe inconsistency, run7 applies simultaneous but misaligned perturbations across the BAO and CMB sectors,  while keeping the SNe dataset fiducial.  This scenario is designed to emulate a realistic multi-probe environment in which BAO favour a slightly higher effective matter density, CMB prefer a lower Hubble constant, and SNe remain internally consistent.  The resulting likelihood contains mutually incompatible ridge directions, providing a controlled example of how multi-probe conflicts can propagate into apparent DDE-like behaviour~\cite{Melchiorri:2006jy,Yang:2021hxg}. 

\subsubsection{Posterior Shifts and Degeneracy Geometry}

The median and $1\sigma$ constraints from the joint MCMC analysis are
\begin{align}
&\Omega_{m0} = 0.3246 \pm 0.0029, \qquad
H_0 = 70.890 \pm 0.167, \nonumber\\[4pt]
&\oo = -1.2054 \pm 0.0162, \qquad
\wa = 1.1102 \pm 0.0238.
\label{run7CP}
\end{align}
Among all runs considered here, run7 produces the largest departure from the fiducial $\Lambda$CDM values $(0.30,70,-1,0)$.  
The inferred matter density is significantly higher than the fiducial model, while $H_0$ shifts moderately above $70$. This counter-intuitive shift compensates for the accelerated expansion driven by the phantom EoS $(\omega < -1)$, which is required to match the CMB acoustic-scale distance while simultaneously satisfying the biased BAO constraints. The CPL parameters move deeply into the phantom-like quadrant $(\oo < -1,\,\wa > 1)$, reflecting the absence of a mutually consistent solution within the combined multi-probe degeneracy structure.

The underlying geometry of this displacement is summarised in Table~\ref{tab:run7_degeneracy}. BAO, CMB, and SNe each prefer incompatible correlations in $(\Omega_{m0},H_0,\oo,\wa)$, so the joint posterior is pushed into a region that satisfies none of the probes individually and is accommodated only through a strongly distorted direction in the CPL parameter space.

\begin{table}[t]
\centering
\caption{Qualitative degeneracy directions relevant for run7. BAO favour higher $\Omega_{m0}$, CMB prefer lower $H_0$, while SNe remain fiducial. The severe misalignment among these directions produces the strongly phantom-like displacement in Eq.~\eqref{run7CP}.  Based on the degeneracy logic described in~\cite{Bernal:2016gxb,Chen:2018dbv}.}
\label{tab:run7_degeneracy}
\vspace{0.2cm}
\begin{tabular}{c|c|c}
\hline
\textbf{Probe} & \textbf{Dominant Degeneracy Direction}
& \textbf{Physical Origin} \\
\hline
BAO (biased: $\Delta\Omega_{m0}>0$) &
$(\Omega_{m0}\uparrow,\ H_0\downarrow)$
& Distorted $D_M/r_d,\ D_H/r_d$ ratios \\[3pt]
CMB (biased: $\Delta H_0<0$) &
$(H_0\downarrow,\ \oo>-1,\ \wa>0)$
& Maintain $(R,\ell_A)$ under low-$H_0$ shift \\[3pt]
SNe (fiducial) &
$(H_0\uparrow,\ \Omega_{m0}\downarrow)$
& Anchored expansion history \\
\hline
\end{tabular}
\end{table}

\subsubsection{Tension-Injection Mechanism}

The multi-probe tension in run7 is generated by combining two independent injections.  First, a positive shift $\Delta\Omega_{m0}>0$ is applied to the BAO model predictions, producing directional distortions in both $D_M(z)/r_d$ and $D_H(z)/r_d$ that move the BAO-only posterior toward larger $\Omega_{m0}$ and mildly lower $H_0$.  Second, a negative shift $\Delta H_0$ is applied to the CMB distance-prior vector $(R,\ell_A)$, mimicking a Planck-like low-$H_0$ preference.  The SNe likelihood is kept completely fiducial, preserving its $(H_0\uparrow,\ \Omega_{m0}\downarrow)$ degeneracy.

The combined likelihood,
\[
\mathcal{L}_{\rm tot}
= \mathcal{L}_{\rm BAO}^{(\Delta\Omega_{m0}>0)}
  \mathcal{L}_{\rm CMB}^{(\Delta H_0<0)}
  \mathcal{L}_{\rm SNe}^{\rm (fid)},
\]
must reconcile three incompatible ridge directions.  Because no consistent solution exists at moderate $(\Omega_{m0},H_0)$, the posterior is driven toward a rapidly evolving and strongly phantom-like EoS, as reflected by the values in Eq.~\eqref{run7CP}. 

\subsubsection{Pivot Equation-of-State}

The pivot quantities for run7 are
\begin{align}
a_p = 0.3846, \qquad
z_p = 1.5999, \qquad
\wpp = -0.5224 \pm 0.0067.
\label{wprun7}
\end{align}
The unusually high pivot redshift ($z_p\simeq1.6$) indicates a strong rotation of the $(\oo,\wa)$ degeneracy direction relative to the other runs. Although the CPL parameters lie in the phantom regime, the pivot value moves into the opposite, quintessence-like region ($\wpp>-1$). Within the CPL basis adopted here, this illustrates how strongly misaligned multi-probe tensions can lead to qualitatively misleading interpretations of late-time DE behaviour.

This large pivot redshift arises from the simultaneous action of mutually inconsistent tension channels in run7. Because BAO, CMB, and SNe impose nearly orthogonal ridge directions in the $(\oo,\wa)$ plane, the combined posterior is forced into a direction that does not align with any individual probe. In the CPL parametrization, the pivot scale satisfies $a_p = 1 + \mathrm{Cov}(\oo,\wa)/\sigma_{\wa}^2$, so a strong negative covariance---produced here by the geometric interference of the three conflicting ridges---drives $a_p$ far below unity and hence pushes $z_p$ to a large value.

This behaviour should not be interpreted as evidence for genuine high-redshift sensitivity to DE. Rather, within the present setup, it reflects the loss of geometric coherence across the probes. The large pivot redshift may therefore be viewed as an indicator of severe multi-probe inconsistency, in the sense that the inferred pivot moves into a regime where none of the probes individually has substantial constraining power.

\subsubsection{Joint Posterior Contours}

\begin{figure}[t]
    \centering
    \includegraphics[width=0.78\linewidth]{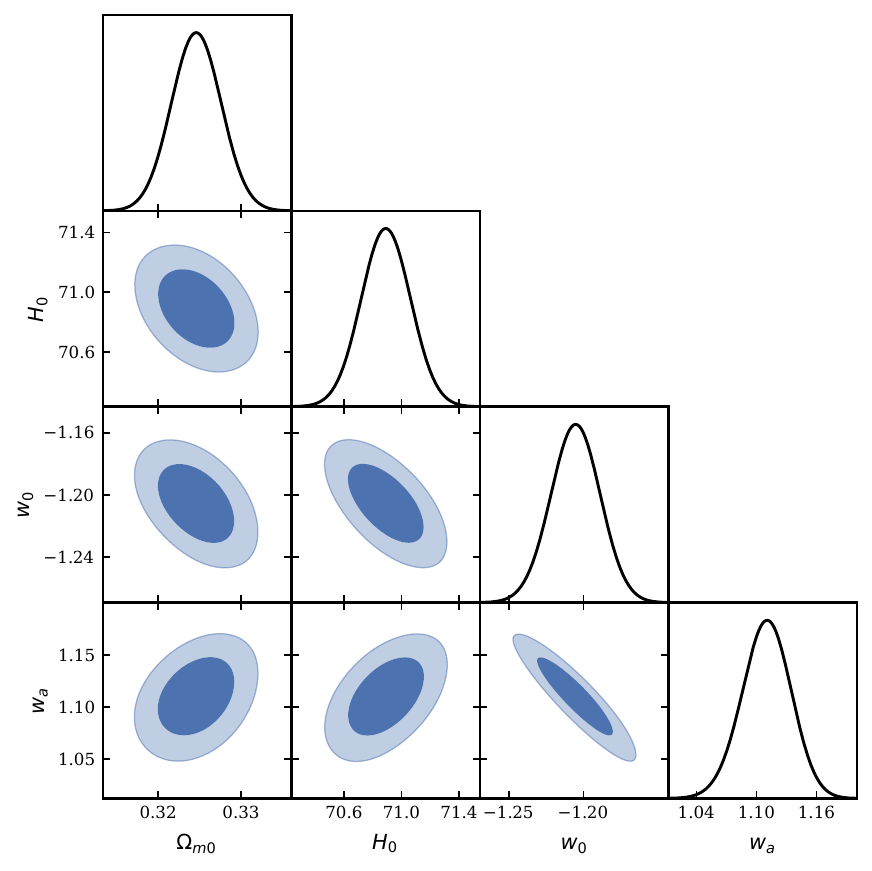}
    \caption{
        \textbf{Run7 (combined multi-probe tension).}
        Joint BAO+CMB+SNe posterior when BAO are biased toward higher 
        $\Omega_{m0}$ and CMB toward lower $H_0$, with SNe fiducial.
        The conflicting ridge directions force the solution into the
        strongly dynamical, phantom-like region:
        $\Omega_{m0}\simeq 0.325$, $H_0\simeq 70.9$,
        $(\oo,\wa)\simeq(-1.21,\,1.11)$.
        The exceptionally large pivot redshift $z_p\simeq1.6$ reflects the 
        extreme rotation of the CPL degeneracy unique to this run.}
    \label{fig:run7_triangle}
\end{figure}

Figure~\ref{fig:run7_triangle} shows the joint posterior for 
$(\Omega_{m0},H_0,\oo,\wa)$.  
The displacement is the largest among all runs examined in this work.  
The $(\Omega_{m0},H_0)$ contours occupy a region that satisfies neither 
the BAO- nor the CMB-favoured direction individually, while the 
$(\oo,\wa)$ contours extend deeply into the 
$(\oo<-1.2,\ \wa>1.1)$ quadrant.  
This confirms that the multi-probe inconsistencies introduced in run7 
produce a qualitative deformation of the CPL parameter space far exceeding 
any of the single-probe tensions previously examined.

\section{Empirical tension--bias transfer functions}
\label{sec:tension_bias_fits}

A central aim of this section is to construct empirical mappings between injected cross-probe tensions and the resulting biases in inferred cosmological parameters.  The Fisher-level expressions introduced earlier provide a useful linear reference for the origin of such mappings.  In practice, however, the full likelihood response is shaped by nonlinear and probe-dependent degeneracy structures, so the calibration of the tension--bias relations must ultimately be obtained from direct MCMC analyses.  Using the controlled tension-injection runs (run2--run7), we therefore construct a set of one-dimensional (1D) and two-dimensional (2D) empirical transfer functions.  The 1D relations describe the local linear response to isolated tension components, whereas the 2D maps capture the coupled effects of $(\Delta\Omega_{m0},\Delta H_0)$ tensions and the nonlinear behaviour that is not visible in the 1D fits.  Taken together, these results provide the empirical basis for the diagnostic forecasting framework developed in this work. These transfer functions should be interpreted in the same restricted sense as the rest of the analysis: they are empirical response summaries within the adopted CPL-plus-compressed-CMB setup.

\subsection{One-Dimensional Tension–Bias Relations}

In this subsection we summarize simple one-dimensional fits that map injected tension parameters onto the resulting posterior biases in $(\Omega_{m0}, H_0, \oo, \wa, \wpp)$, based on the controlled tension-injection runs (run2--run7).

\subsubsection{Tension in $\Omega_{m0}$}

In this scenario we inject a controlled offset $\Delta\Omega_{m0}$ and measure the resulting parameter shifts
\begin{align}
  & \Delta\Omega_{m0}^{\rm (post)} \simeq 0.9749\,\Delta\Omega_{m0} + 1.0170\times 10^{-2}, 
  \Delta H_0^{\rm (post)} \simeq -1.1247\times 10^{1}\,\Delta\Omega_{m0} + 6.0557\times 10^{-1},\nonumber \\
  &\Delta \oo \simeq 0.6533\,\Delta\Omega_{m0} - 1.7934\times 10^{-1}, \qquad
  \Delta \wa \simeq 7.1200\,\Delta\Omega_{m0} + 7.5432\times 10^{-1}.
      \label{eq:fit_dOm0}
\end{align}

\subsubsection{Tension in $H_0$}
Injecting a controlled offset $\Delta H_0$ produces
\begin{align}
  &\Delta\Omega_{m0}^{\rm (post)} \simeq -3.6935\times 10^{-3}\,\Delta H_0 + 1.8467\times 10^{-2},
  \Delta H_0^{\rm (post)} \simeq -2.6402\times 10^{-2}\,\Delta H_0 + 1.3201\times 10^{-1}, \nonumber \\
  &\Delta \oo \simeq 1.6512\times 10^{-2}\,\Delta H_0 - 8.2562\times 10^{-2}, \qquad
  \Delta \wa \simeq -9.3891\times 10^{-2}\,\Delta H_0 + 4.6945\times 10^{-1}. \label{eq:fit_dH0}
\end{align}

\subsubsection{SNe absolute-magnitude tension}

Injecting a SNe absolute-magnitude shift $\Delta M$ yields
\begin{align}
  &\Delta\Omega_{m0}^{\rm (post)} \simeq -6.9189\times 10^{-1}\,\Delta M - 3.0364\times 10^{-2},
  \Delta H_0^{\rm (post)} \simeq 7.8093\times 10^{1}\,\Delta M + 3.5290,\nonumber \\ 
 &\Delta \oo \simeq -2.1823\,\Delta M - 9.2155\times 10^{-2}, \qquad
  \Delta \wa \simeq -6.8433\times 10^{-1}\,\Delta M - 5.4823\times 10^{-2}.
      \label{eq:fit_dM}
\end{align}

\subsection{From One-Dimensional Fits to a Multi-Dimensional Tension--Bias Map}
\label{subsec:multid_map}

The 1D relations above describe the response to single and isolated tension components.  However, realistic analyses—such as DESI DR2 combined with Planck and SNe—generally involve multiple, correlated tensions.  To accommodate such cases, we extend the 1D results into a multi-dimensional linear map.
\begin{equation}
\Delta\boldsymbol{\theta}_{\rm post}
\simeq
\mathbf{A}\,\Delta\boldsymbol{t},
\quad
\Delta\boldsymbol{t}
=(\Delta\Omega_{m0},\Delta H_0,\Delta M)^{\mathsf T},\quad
\textrm{with} \quad
\Delta\boldsymbol{\theta}_{\rm post}
=(\Delta\Omega_{m0}^{\rm post},
  \Delta H_0^{\rm post},
  \Delta \oo,
  \Delta \wa)^{\mathsf T}. \label{eqDeltatheta}
\end{equation}

\subsubsection{Construction of the Bias Matrix}

Using the slopes in Eqs.~\eqref{eq:fit_dOm0}--\eqref{eq:fit_dM}, we obtain the $4\times 3$ empirical bias matrix
\begin{equation}
\mathbf{A}=
\begin{pmatrix}
0.97493   & -3.6935\times10^{-3} & -0.69189 \\
-11.247   & -2.6402\times10^{-2} & 78.093   \\
0.65334   &  1.6512\times10^{-2} & -2.1823  \\
7.1200    & -9.3891\times10^{-2} & -0.68433
\end{pmatrix}.
\label{eq:Amatrix}
\end{equation}

\subsubsection{Two-Dimensional Bias Maps and Vector Diagrams}

The linear relations summarized above characterize only the local, independent responses to single injected tensions.  However, when multiple tension components are present simultaneously—as is common in DESI DR2, Planck, and SNe combinations—the resulting posterior biases in $(\oo,\,\wa)$ need not follow the simple 1D slopes.  In particular, $\Delta\Omega_{m0}$ and $\Delta H_0$ jointly distort the BAO and CMB likelihoods in a non-separable way, producing curved and anisotropic structures in the inferred DE parameters.  To illustrate these multidimensional effects, we construct empirical 2D tension–bias maps for
$(\Delta\oo,\Delta\wa)$ using the full set of run2–run7 simulations.

\begin{figure}[t]
\centering
\begin{minipage}[t]{0.48\linewidth}
    \centering
    \includegraphics[width=\linewidth]{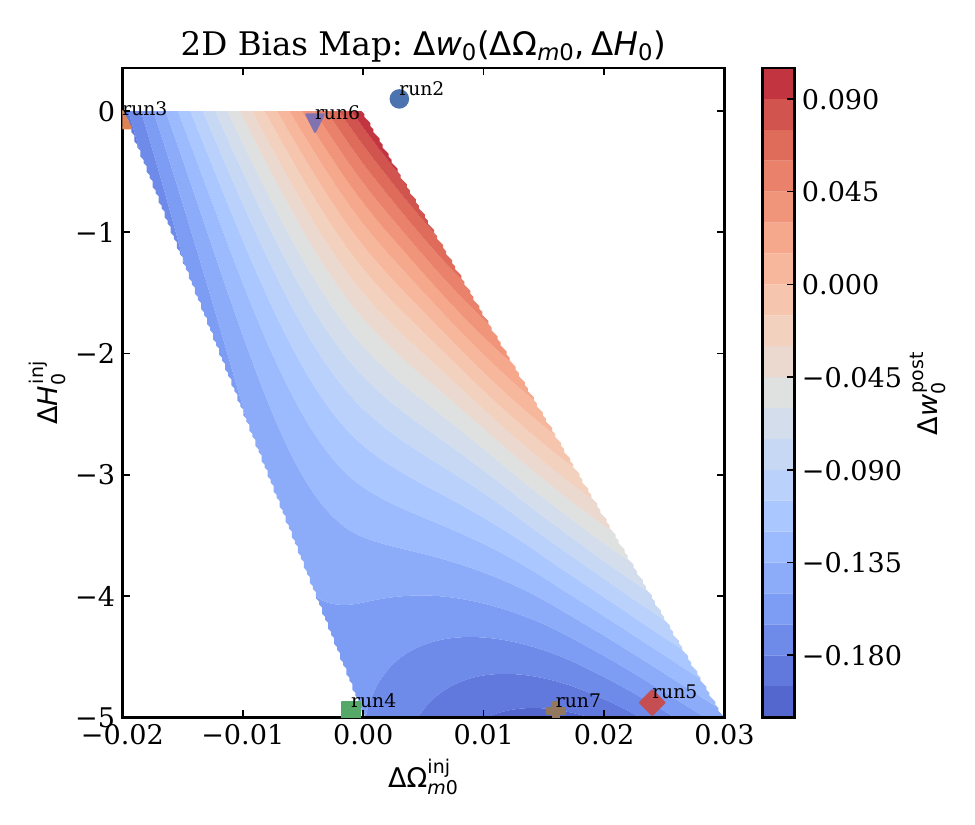}
    \vspace{2pt}
    \textbf{(a) $\Delta\oo$ map}
    \label{fig:2dmap_w0}
\end{minipage}
\hfill
\begin{minipage}[t]{0.48\linewidth}
    \centering
    \includegraphics[width=\linewidth]{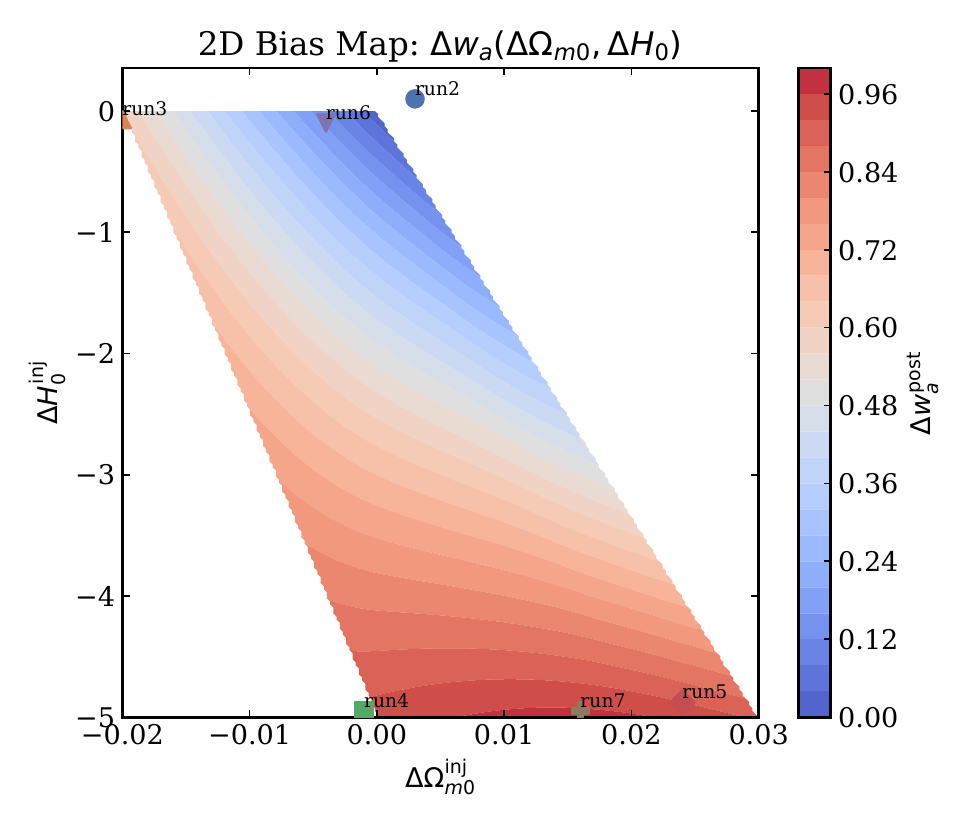}
    \vspace{2pt}
    \textbf{(b) $\Delta\wa$ map}
    \label{fig:2dmap_wa}
\end{minipage}
\caption{Two-dimensional tension--bias maps for $\Delta\oo$ and $\Delta\wa$ as functions of the injected tensions $(\Delta\Omega_{m0},\Delta H_0)$, constructed from run2--run7.  Both panels show significant curvature and anisotropy, indicating nonlinear mixing of the tension directions even in the small-tension regime.}
\label{fig:2d_bias_maps}
\end{figure}

Figure~\ref{fig:2d_bias_maps} shows the two empirical 2D tension–bias maps constructed directly from the run2–run7 sampling.  
Since each run corresponds to a distinct direction in the$(\Delta\Omega_{m0},\Delta H_0)$ plane, the morphology of the two maps is largely determined by how these six runs populate and distort the local neighbourhood around the origin. In the left panel, the curvature of $\Delta\oo(\Delta\Omega_{m0},\Delta H_0)$ arises from the interplay between BAO-oriented runs (run3 and run5), which shift the map along direction with shallow $H_0$ variation, and $H_0$-dominated runs (run2 and run6), which steepen the vertical gradient.  This superposition produces the diagonal warping seen in the contour lines.  The combined effect produces the diagonal warping visible in the contour lines: the BAO-side runs populate the lower diagonal half of the map, whereas the CMB/SNe-like runs dominate the upper vertical region,leading to the characteristic deformation toward the $D_H/r_d$ direction.

In the right panel, the map of $\Delta\wa$ shows an even more nonlinear response.  The pronounced curved ridge originates from the strong leverage of run2 and run6, both of which induce substantial effective shifts in the CMB acoustic angular scale and project almost entirely onto the steep $(\oo,\wa)$ degeneracy direction. Runs with primarily BAO-side tensions (run3 and run5) populate the left–right diagonal portion of the map, stretching the ridge along a direction consistent with the BAO constraint surface in the $(\Omega_{m0},H_0)$ plane.  Because run7 combines BAO, CMB, and SNe tensions simultaneously, its influence appears as a smooth interpolation between these directions,further thickening the ridge and producing the nonlinear twist seen in the central region of the map.

Taken together, the two panels show that the full 2D structure of the tension–bias relationship arises from the superposition of the discrete run2–run7 tension directions.  The BAO-dominated runs imprint a diagonal shear, the CMB/SNe-dominated runs imprint a nearly vertical distortion, and the mixed-tension run7 provides the curved interpolation between these axes.  This explains why the resulting maps exhibit pronounced curvature and ridge-like anisotropies even in the nominally small-tension regime.

\begin{figure}[t]
    \centering
    \includegraphics[width=0.70\linewidth]{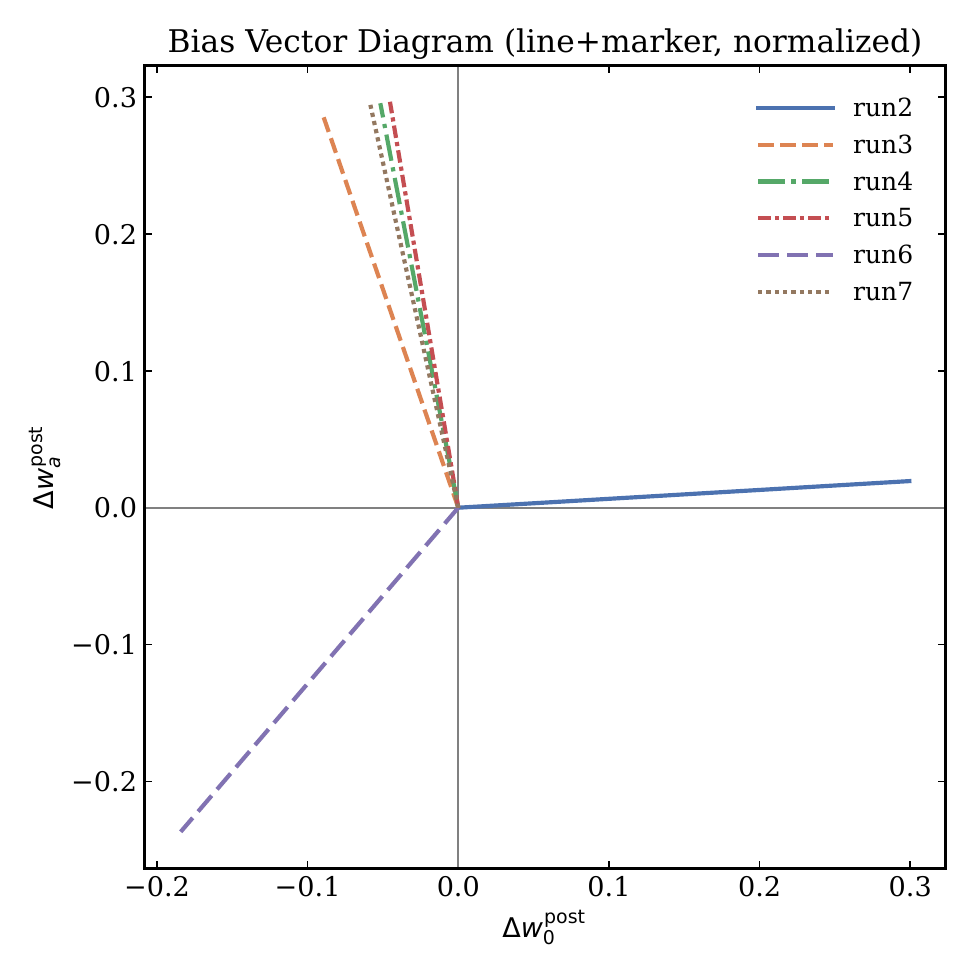}
    \caption{Normalized bias vectors in the $(\oo,\wa)$ plane for runs~2--7, plotted as line segments of unit length.  
Each segment originates at the origin and terminates at the normalized direction $(\Delta\oo,\Delta\wa)/\|(\Delta\oo,\Delta\wa)\|$.  
Runs are distinguished by colour and line style, as indicated in the legend.  The clustering of the segments along a small number of preferred orientations reflects the dominant roles of BAO radial-distance constraints and CMB acoustic-scale geometry in shaping tension-driven distortions in DE parameter space.}
    \label{fig:bias_vectors}
\end{figure}

While the 2D maps reveal the local behaviour across the tension plane, the global distortion pattern becomes clearer when the responses are represented as normalized line segments in the $(\oo,\wa)$ plane, as shown in Fig.~\ref{fig:bias_vectors}.  
Each run contributes a directional line segment starting at the origin and ending at the normalized point $(\Delta\oo,\Delta\wa)/\|(\Delta\oo,\Delta\wa)\|$, with the individual runs distinguished by colour and line style, as indicated in the legend.  BAO--dominated runs (such as run3 and run5) produce line segments that lie nearly orthogonal to the CMB acoustic-scale degeneracy, whereas $H_0$--dominated runs (run2 and run6) are aligned with the steep CMB-preferred ridge, resulting in comparatively larger excursions in $\Delta\wa$.  Run7 lies between these behaviours, reflecting the mild nonlinearity already visible in the 2D tension--bias maps. These patterns demonstrate that tension-induced distortions in DE parameter space are highly directional: cosmological constraints are intrinsically more sensitive to specific linear combinations of $(\Delta\Omega_{m0},\Delta H_0)$ than to either tension alone.

These combined results demonstrate that even modest cross-probe tensions generate a highly anisotropic response in $(\oo,\wa)$, with distinct ridge directions and nonlinear mixing between $\Delta\Omega_{m0}$ and $\Delta H_0$.  This behaviour directly motivates the multidimensional analysis in Sec.~\ref{subsec:limits_1D} and explains why the 1D transfer functions systematically fail for large, realistic tensions such as those appearing in DESI\,DR2\,+\,CMB.

\subsubsection{Predicting Multi-Probe Biases}

For any hypothetical tension vector $\Delta\boldsymbol{t}$, the predicted posterior shift follows directly from $\Delta\boldsymbol{\theta}_{\rm post}=\mathbf{A}\Delta\boldsymbol{t}$.  
Conversely, given an observed shift $\Delta\boldsymbol{\theta}_{\rm obs}$, the effective tension vector is
\begin{equation}
\Delta\boldsymbol{t}_{\rm eff}
\simeq
(\mathbf{A}^{\mathsf T}\mathbf{A})^{-1}
\mathbf{A}^{\mathsf T}
\Delta\boldsymbol{\theta}_{\rm obs}.
\label{eq:inverse_tension}
\end{equation}

\subsection{Limitations of the One-Dimensional Transfer Functions}
\label{subsec:limits_1D}

The one-dimensional tension--bias relations derived in the previous subsection offer a transparent and intuitive picture of how isolated tension components propagate into shifts of $(\oo,\wa)$.  Within the present analysis, however, these relations should be understood as local empirical diagnostics calibrated in the CPL parametrization, rather than as universal mappings between probe-level inconsistencies and dark-energy evolution. These relations are intrinsically local: they capture only the linear response of the posterior around the fiducial model and within the modest tension amplitudes explored in run2--run7.  When applied to large or highly correlated tensions such as those observed in the DESI\,DR2\,+\,CMB combination, the limitations of the 1D framework become immediately apparent.

To illustrate this breakdown concretely, consider the DESI\,DR2\,+\,CMB deviations from the fiducial $(0.30,70,-1,0)$,
\[
\Delta\boldsymbol{\theta}_{\rm obs}
\simeq
(+0.052,\,-6.3,\,+0.57,\,-1.70)^{\mathsf T}.
\]
Using the empirical linear map of Eq.~\eqref{eq:inverse_tension}, we may infer the effective tensions required to reproduce these shifts.  
The resulting value,
\[
\Delta\Omega_{m0}^{\rm inj}\simeq +0.044,
\qquad
\Delta H_0^{\rm inj}\simeq -5.2~{\rm km\,s^{-1}Mpc^{-1}},
\qquad
\Delta M^{\rm inj}\approx 0,
\]
correctly identifies the direction of the DESI--CMB discrepancy: a BAO-driven enhancement of $\Omega_{m0}$ combined with a CMB-preferred suppression of $H_0$.  
Yet, when these inferred tensions are mapped back to the DE sector using the 1D relations, the predicted biases,
\[
\Delta\oo^{\rm pred}\approx -0.057,
\qquad
\Delta\wa^{\rm pred}\approx -0.175,
\]
fall short by nearly an order of magnitude compared with the observed
\[
\Delta\oo^{\rm obs}\simeq +0.57, \qquad 
\Delta\wa^{\rm obs}\simeq -1.70.
\]

This discrepancy is not a flaw of the method but rather a reflection of the fact that the 1D transfer functions were never designed to operate far from the fiducial region.  
Three effects underlie their breakdown.  
First, the calibration range of run2--run7 intentionally restricts the injected tensions to modest amplitudes, ensuring numerical stability and maintaining a locally linear response.  
Second, the CMB acoustic-scale likelihood possesses a strongly curved degeneracy direction in $(\oo,\wa)$: once the acoustic scale is displaced beyond the local regime, the curvature of this degeneracy ridge becomes dominant and the response ceases to be linear in $(\Delta\Omega_{m0},\Delta H_0)$.  
Third, real CMB-side tensions are accompanied by correlated shifts in $(r_d,\omega_b,n_s)$, none of which are included in the minimal three-parameter tension vector $(\Delta\Omega_{m0},\Delta H_0,\Delta M)$.  
The 1D framework cannot reproduce these coupled effects, and therefore it naturally underestimates large, nonlinear displacements.

A related point is that the detailed form of the 1D mappings depends not only on the locality of the expansion but also on the adopted likelihood representation. In particular, the reduced CMB description used in this work compresses the full CMB information into a geometric subspace and models that reduced space in Gaussian form. This is adequate for the controlled forecasting purpose of the present analysis, but it can modify the effective degeneracy widths and orientations relative to a fuller likelihood treatment once the posterior is driven far from the fiducial region.

Nevertheless, despite their limitations beyond the local regime, the 1D transfer functions remain useful within the controlled-tension forecasting framework developed here.  
They provide an immediate and interpretable link between specific probe-level tensions and the resulting shifts in cosmological parameters, allowing one to identify which sector---BAO, CMB, or SNe---dominates the observed bias.  
More importantly, they offer a diagnostic baseline against which the fully nonlinear behaviour seen in 2D tension--bias maps or in real-data combinations can be quantified.  
The failure of the 1D approximation at large tensions is itself informative: it signals the onset of nonlinear mixing, identifies where curved degeneracy directions begin to dominate, and highlights the need for higher-order extensions such as global quadratic models, Fisher-bias corrections including covariance derivatives, or nonparametric Gaussian-process emulators trained on extended injection grids.

In addition, the numerical values of these transfer functions should not be interpreted as parametrization-independent. A different late-time basis or a more flexible dark-energy description would generally alter both the slope and curvature of the inferred tension--bias relation, even if the broader conclusion---that cross-probe inconsistencies can be absorbed by a flexible late-time sector---remains qualitatively similar.

In this sense, the 1D relations are not a substitute for a full multi-dimensional treatment.  
Instead, they provide a useful local diagnostic reference.  They characterize the local structure of tension propagation, help diagnose departures from linear behaviour in real data, and guide the development of higher-order or emulator-based extensions of the tension to bias mapping.

\section{Discussion}
\label{sec:discussion}

The controlled tension--injection framework developed in this work illustrates how apparent hints of DDE can arise even when the underlying cosmology is exactly $\Lambda$CDM. By explicitly separating the origin of the imposed bias---the injected tension vector $\Delta\mathbf{t}$---from the statistical response of each probe, the framework provides a controlled way to interpret multi-probe inconsistencies that are otherwise difficult to disentangle. The results should, however, be understood within the specific scope of the present setup: a CPL-based analysis with compressed CMB priors and deliberately simplified tension injections.

Within this setting, our simulations highlight three recurring features of cross-probe tensions. First, BAO, CMB, and SNe respond differently to the same physical perturbation because they constrain distinct combinations of absolute and relative distance scales. A single tension such as $\Delta H_0$ induces nearly orthogonal shifts in the individual probes, and their intersections yield partial compensation rather than full cancellation. Residual misalignments in $(\Omega_{m0},H_0)$ then project onto the steep degeneracy directions of $(\oo,\wa)$, generating posterior shifts that can resemble DDE-like behaviour.

Second, Fisher-level linear predictions successfully reproduce the direction of these shifts across all scenarios, even when they underestimate the amplitude for larger tensions. In this sense, the Fisher analysis is useful mainly as a local diagnostic of the response around the fiducial model. Its success in tracing the direction of the bias does not imply that the full posterior response remains linear once the injected inconsistency becomes large or multiple tension channels are combined.

Third, modest and observationally motivated tensions are already sufficient to displace the joint BAO+CMB+SNe posterior away from $(\oo,\wa)=(-1,0)$ at the $1$--$2\sigma$ level. This result should not be read as a direct statement about any particular real-data combination. Rather, it shows that in a controlled mock setting, relatively small inter-probe mismatches can be projected into the CPL dark-energy sector if the probe geometries are not fully aligned. This highlights the importance of carefully synchronizing the absolute calibrations implicit in different probes before interpreting any deviation from $\Lambda$CDM as physical.

An instructive outcome appears in the multi-source tension scenario (run7). The CPL parameters $(\oo,\wa)$ move into the phantom regime, yet the pivot parameter shifts to $\wpp \simeq -0.52$ at an anomalously high pivot redshift $z_p \simeq 1.6$. This occurs because strongly misaligned tension vectors rotate the CPL degeneracy direction far beyond the region of direct data leverage. Within the CPL basis, this example shows that the pivot interpretation can become qualitatively misleading when the underlying probe combination is strongly inconsistent.

At the same time, the present results are not expected to be parametrization-independent in a strict sense. The detailed amplitudes and directions of the inferred shifts in dark-energy parameter space depend on the choice of late-time basis, and CPL is known to be particularly sensitive to degeneracy geometry and prior structure. The main lesson of this work is therefore not that the specific transfer functions reported here are universal, but that cross-probe inconsistencies can be absorbed by a flexible late-time sector in a structured and potentially misleading way. If one moves beyond the CPL parametrization, the detailed response found here would in general change. For example, alternative two-parameter late-time bases, binned or principal-component descriptions of $\omega(z)$, or physically restricted dark-energy models could rotate the degeneracy directions, redistribute the apparent bias between amplitude and evolution, or reduce the tendency of the fit to express the tension specifically through the CPL pair $(\omega_0,\omega_a)$. In that sense, the patterns reported here should be read as basis-dependent diagnostic responses of a widely used phenomenological parametrization, rather than as unique signatures that would necessarily persist unchanged in a more general late-time description.

A related limitation concerns the CMB treatment. In this work, the CMB sector is represented by compressed distance priors rather than by a full CMB likelihood. This choice is useful for isolating the geometric contribution of the CMB in a transparent forecasting setup, but it necessarily suppresses correlations with additional parameters such as $n_s$, $A_s$, and $\tau$. As a result, the effective degeneracy structure in the reduced space should be interpreted as an approximation to the full problem, not as a complete surrogate for end-to-end Planck-like analyses. This caution is further supported by recent studies showing that the inferred degeneracy structure in dynamical dark-energy analyses can remain sensitive to additional degrees of freedom, such as the CMB lensing amplitude~\cite{Park:2025fbl}.

Likewise, the injected tensions considered here are intentionally simple. They are implemented mainly as coherent shifts in a small number of control parameters, whereas realistic observational systematics can be redshift dependent, observable dependent, and correlated across probes. The current setup is therefore best viewed as a controlled first step that isolates leading projection effects, rather than as a full model of survey systematics. Extensions to more realistic injection patterns, including redshift-dependent distortions and fuller covariance-level treatments, would be a natural next step.

With these qualifications in mind, the injection framework still provides a useful way to anticipate how future analyses may respond to calibration shifts, inconsistent priors, or mild cross-probe mismatches. Its main value is diagnostic: it provides a reproducible mock-based setting in which the geometric response of multi-probe inference can be stress-tested before stronger physical conclusions are drawn from apparent departures from $\Lambda$CDM.

\section{Conclusion}
\label{sec:conclusion}

We have presented a controlled tension--injection framework for quantifying how probe--probe inconsistencies can propagate into inferred dark-energy parameters within a simplified mock-based setting. By parameterizing departures from mutual consistency in the BAO, CMB, and SNe sectors and tracing their impact through full MCMC analyses, the method separates the physical origin of bias (the tension vector $\Delta\mathbf{t}$) from the statistical response of $(\Omega_{m0},H_0,\oo,\wa)$.

Within the CPL-based framework considered here, our results show three main features. First, even modest probe-level tensions can displace the joint posterior away from $(\oo,\wa)=(-1,0)$, despite the underlying cosmology being exactly $\Lambda$CDM. Second, the direction of the resulting shift depends not only on the injected tension itself but also on the relative degeneracy geometry of BAO, CMB, and SNe, so that the final multi-probe posterior need not follow the naive single-probe trend. Third, the empirical 1D tension--bias relations provide useful local diagnostics, while the 2D maps reveal nonlinear mixing between tension components and identify the combinations of $(\Delta\Omega_{m0},\Delta H_0)$ to which the inferred dark-energy constraints are most sensitive.

In particular, the controlled runs show that SNe-side calibration shifts, BAO-side matter-density distortions, and CMB-side acoustic-scale mismatches project into qualitatively different regions of the $(\oo,\wa)$ plane, while combined multi-probe inconsistencies can produce more extreme and potentially misleading behaviour, such as the anomalously high pivot redshift found in run7. These results illustrate how apparent dynamical-dark-energy-like signatures can emerge from cross-probe inconsistency even in a mock universe generated from fiducial $\Lambda$CDM.

These findings underscore the importance of interpreting modern multi-probe analyses in light of how different datasets encode absolute and relative distance scales. Within the restricted setup adopted here, the framework developed in this work is best understood as an illustrative and diagnostic tool: it provides a controlled mock-based route to studying apparent dynamical-dark-energy-like signatures, assessing the robustness of joint-probe constraints, and identifying which sectors of an analysis pipeline are most susceptible to tension-induced bias.

At the same time, the present results should not be interpreted as parametrization-independent, nor as a substitute for full-likelihood analyses of current data. Their quantitative details depend on the CPL basis, on the compressed CMB representation adopted here, on the approximate Gaussian treatment of the reduced CMB prior, and on the deliberately simplified form of the injected tensions. The main value of the present analysis is therefore to clarify how controlled inconsistencies can propagate geometrically through a common inference pipeline, rather than to provide a fully realistic prediction of the magnitude of such effects in any specific observational dataset.

Future work may extend this framework to broader late-time parametrizations, early-dark-energy or varying-sound-horizon models, and modified-gravity scenarios, as well as to more realistic injection patterns and fuller likelihood treatments. In that broader context, the controlled approach developed here may serve as a useful baseline for separating genuine physical signals from artefacts associated with subtle cross-probe inconsistencies.

\appendix
\section{Technical Supplement: Controlled Tension Injection and Fisher-Level Bias Propagation}
\label{sec:appendix_technical}

This appendix summarizes the essential theoretical and methodological elements underlying the controlled tension–injection framework developed in the main text.  Our aim is to provide a compact reference for the analytic structure of parameter biases, the construction of tension-modified likelihoods, and the interpretation of BAO, CMB, and SNe observables within Fisher theory.  The material below complements, but does not repeat, the main discussion.

\subsection{Science Rationale and Conceptual Goals}
\label{sec:science_goals}

The controlled tension–injection project is designed to create a fully transparent environment in which cross-probe inconsistencies can be introduced and their impact on cosmological inference quantified.  
Within this framework, well-defined perturbations in parameters such as $\Omega_{m0}$, $H_0$, the SNe absolute magnitude $M$, and the BAO sound horizon $r_d$ are applied to otherwise self-consistent mocks.  
The resulting parameter shifts after running full likelihood analyses allow us to empirically map 
\begin{align}
\Delta\mathbf{t}
\longrightarrow
\Delta(\oo, \wa, \wpp,\Omega_{m0},H_0), \label{AppDeltat}
\end{align}
where $\Delta\mathbf{t}$ denotes the injected tension parameters. In the main text, these mappings are used as empirical tension--bias relations within the restricted setting adopted there.

\subsection{Design of the Controlled Tension Pipeline}
\label{sec:pipeline_design}

Tensions are introduced by modifying selected components of the data vector, priors, or mock generation procedure relative to a common fiducial $\Lambda$CDM cosmology.  
Typical cases include: (i) shifts in $\Omega_{m0}$ motivated by BAO–SNe inconsistencies;  
(ii) $H_0$ offsets mimicking SH0ES–Planck tension;  
(iii) zero-point or redshift-dependent calibration drifts in SNe; and  
(iv) changes in $r_d$ or its effective value for BAO.  
Each tension is parametrized by control variables such as $\Delta\Omega_{m0}$, $\Delta H_0$, or $\Delta M$, which can be scanned over a grid.  For each injected configuration, MCMC analyses of individual or joint probe combinations are performed, and the resulting posterior shifts define the empirical tension–bias relations presented in the main text.

\subsection{Fisher-Level Bias from SNe Calibration Drift}
\label{sec:fisher_bias_tilt}

A subtle but important source of cross-probe tension arises from redshift-dependent evolution in the SNe~Ia calibration.  
If the true absolute magnitude varies as  
\begin{align}
M_{\rm true}(z) = M_0 + \Delta M_0 + \Delta M_1 z, \label{AppM}
\end{align}
then the term proportional to $z$ cannot be absorbed by a constant nuisance parameter $M$ and inevitably induces biases in $(\oo,\,\wa)$.  
The resulting systematic residual in the Hubble diagram is
\begin{align}
\Delta\mu(z_i) = \Delta M_0 + \Delta M_1 z_i,
\label{AppDeltamu}
\end{align}
and the linear tilt term sources a bias vector
\begin{align}
b_\alpha^{(1)} = \Delta M_1 \sum_{i,j}
\frac{\partial\mu(z_i)}{\partial\theta_\alpha}
(C^{-1})_{ij} \, z_j,
\label{Appbalpha1} 
\end{align}
for parameters $\theta_\alpha\in\{\Omega_{m0},H_0,\oo,\wa,M\}$.  
The induced cosmological biases follow the standard Fisher expression,
\begin{align}
\Delta\theta_\alpha
=
\sum_\beta (F^{-1})_{\alpha\beta} b_\beta, \label{AppDeltatheta}
\end{align}
with the tilt component projecting most strongly onto $(\oo,\wa)$. This expression makes explicit how a redshift-dependent calibration term can produce an apparent evolution in the inferred dark-energy sector within the adopted fitting basis.

\subsection{Fisher-Level Bias from $\Delta\Omega_{m0}$, $\Delta H_0$, and $\Delta r_d$}
\label{sec:fisher_bias_standard}

Tensions in $\Omega_{m0}$, $H_0$, and $r_d$ arise frequently in joint analyses of DESI BAO, SNe, and CMB distance priors.  
If the theoretical model is displaced relative to the true cosmology by $\Omega_{m0}\rightarrow\Omega_{m0}+\Delta\Omega_{m0}, H_0\rightarrow H_0+\Delta H_0,r_d\rightarrow r_d+\Delta r_d$,  then the induced residual in any probe’s observable $d_i$ takes the linearized form
\begin{align}
\Delta d_i=
-\frac{\partial d_i}{\partial\Omega_{m0}}\Delta\Omega_{m0}
-\frac{\partial d_i}{\partial H_0}\Delta H_0
-\frac{\partial d_i}{\partial r_d}\Delta r_d.
\label{AppDeltadi}
\end{align}
The corresponding parameter bias is again determined by the Fisher–bias relation,
\begin{align}
\Delta\theta_\alpha = -\sum_{\beta}(F^{-1})_{\alpha\beta} \Big[ \Delta\Omega_{m0} S_\beta^{(\Omega)} +
\Delta H_0 S_\beta^{(H)} + \Delta r_d S_\beta^{(r_d)} \Big], \label{AppDeltathetaalpha}
\end{align}
where the projection coefficients $S_\beta^{(X)}$ encode how each probe transmits a given tension into the CPL parameter space. These expressions provide a linearized reference for interpreting the empirical bias relations shown in the main text.

\subsection{Analytic Derivatives for DESI BAO and CMB Distance Priors}
\label{sec:analytic_derivatives}

For completeness, we summarize the derivatives entering the Fisher-level expressions.  
The DESI BAO observables,$d_1(z)=D_M(z)/r_d, d_2(z)=D_H(z)/r_d$, yield the simple $r_d$ derivative  
\begin{align}
\frac{\partial d_{1,2}}{\partial r_d}=-\frac{d_{1,2}}{r_d}, \label{Appd12} 
\end{align}
while derivatives with respect to $\Omega_{m0}$ involve  
$\partial H(z)/\partial\Omega_{m0}$ and its effect on $D_M(z)$ through line-of-sight integration.  
Likewise, for the CMB distance priors $(R,\ell_A)$, the relevant derivatives combine $\partial D_M(z_*)/\partial\theta$ and $\partial r_s(z_*)/\partial\theta$, with their propagation into
\begin{align}
R = \sqrt{\Omega_{m0}}\,H_0\,D_M(z_*)/c, 
\qquad
\ell_A = \pi D_M(z_*)/r_s(z_*). \label{AppRla}
\end{align}
These derivatives determine how perturbations in early-time or late-time parameters shift the acoustic scale and distance ratio constraints in CMB+BAO joint analyses.

\subsection{Forecasting Workflow}
\label{sec:forecasting_workflow}

The workflow adopted in this paper proceeds by generating fiducial BAO, CMB, and SNe mocks, introducing controlled tensions in selected parameters, running MCMC analyses on individual and combined probes, and measuring the resulting parameter shifts. The displacements in $(\Omega_{m0},H_0,\oo,\wa,\wpp)$ as functions of the injected tensions define the empirical relations used in the main text.  Within the limitations discussed in the main text, these relations provide only a compact diagnostic summary of how controlled inconsistencies propagate through the chosen inference pipeline.

\begin{acknowledgments}
This work is supported by Basic Science Research Program through the National Research Foundation of Korea
(NRF) funded by the Ministry of Science and ICT under the Grants No. NRF-RS-2021-NR059413. 
\end{acknowledgments}



\end{document}